\documentclass[sigconf]{acmart}

\graphicspath{ {./figures/} }

\usepackage{listings}
\usepackage{multirow}

\AtBeginDocument{%
  \providecommand\BibTeX{{%
    \normalfont B\kern-0.5em{\scshape i\kern-0.25em b}\kern-0.8em\TeX}}}

\copyrightyear{2024}
\acmYear{2024}
\setcopyright{rightsretained}
\acmConference[IUI '24]{29th International Conference on Intelligent User Interfaces}{March 18--21, 2024}{Greenville, SC, USA}
\acmBooktitle{29th International Conference on Intelligent User Interfaces (IUI '24), March 18--21, 2024, Greenville, SC, USA}
\acmDOI{10.1145/3640543.3645164}
\acmISBN{979-8-4007-0508-3/24/03}

\begin{document}

\title{ExpressEdit: Video Editing with Natural Language and Sketching}

\newcommand{\sysname}[0]{ExpressEdit}

%

\author{Bekzat Tilekbay}
\email{tilekbay@kaist.ac.kr}
\orcid{0009-0003-0002-5400}
\affiliation{%
  \institution{School of Computing, KAIST}
  \city{Daejeon}
  \country{Republic of Korea}
}

\author{Saelyne Yang}
\email{saelyne@kaist.ac.kr}
\orcid{0000-0003-1776-4712}
\affiliation{%
  \institution{School of Computing, KAIST}
  \city{Daejeon}
  \country{Republic of Korea}
}

\author{Michal Lewkowicz}
\email{michal.lewkowicz@yale.edu}
\orcid{0009-0008-3522-2259}
\affiliation{%
  \institution{Department of Computer Science, Yale University}
  \city{New Haven, Connecticut}
  \country{USA}
}

\author{Alex Suryapranata}
\email{alextio@kaist.ac.kr}
\orcid{0009-0008-9494-2256}
\affiliation{%
  \institution{School of Computing, KAIST}
  \city{Daejeon}
  \country{Republic of Korea}
}

\author{Juho Kim}
\email{juhokim@kaist.ac.kr}
\orcid{0000-0001-6348-4127}
\affiliation{%
  \institution{School of Computing, KAIST}
  \city{Daejeon}
  \country{Republic of Korea}
}



\begin{abstract}

Informational videos serve as a crucial source for explaining conceptual and procedural knowledge to novices and experts alike. When producing informational videos, editors edit videos by overlaying text/images or trimming footage to enhance the video quality and make it more engaging. However, video editing can be difficult and time-consuming, especially for novice video editors who often struggle with expressing and implementing their editing ideas. To address this challenge, we first explored how multimodality---natural language (NL) and sketching, which are natural modalities humans use for expression---can be utilized to support video editors in expressing video editing ideas. We gathered 176 multimodal expressions of editing commands from 10 video editors, which revealed the patterns of use of NL and sketching in describing edit intents. Based on the findings, we present \sysname{}, a system that enables editing videos via NL text and sketching on the video frame. Powered by LLM and vision models, the system interprets (1) temporal, (2) spatial, and (3) operational references in an NL command and spatial references from sketching. The system implements the interpreted edits, which then the user can iterate on. An observational study (N=10) showed that \sysname{} enhanced the ability of novice video editors to express and implement their edit ideas. The system allowed participants to perform edits more efficiently and generate more ideas by generating edits based on user's multimodal edit commands and supporting iterations on the editing commands. This work offers insights into the design of future multimodal interfaces and AI-based pipelines for video editing.
 
\end{abstract}

\ccsdesc[500]{Human-centered computing~Interaction paradigms}
\ccsdesc[300]{Human-centered computing~Interactive systems and tools}

\keywords{video editing, human-AI interaction, multimodal input}

\setlength{\belowcaptionskip}{10pt}

\begin{teaserfigure}
  \includegraphics[width=\textwidth]{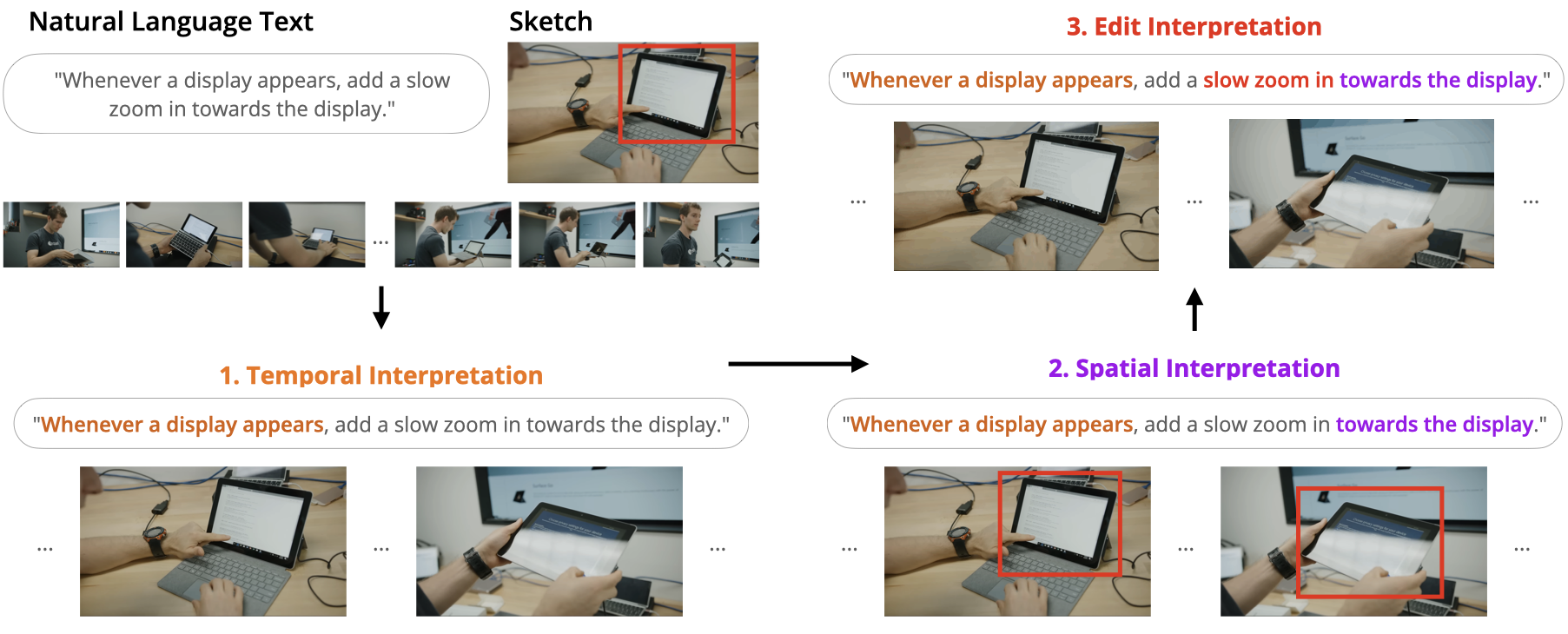}
  \caption{\sysname{} is a multimodal video editing system that supports implementing edits by interpreting temporal, spatial, and edit operation and parameter references in the NL text and sketching edit command.}
    \Description{The figure shows the overall flow of the processing of the multimodal (natural language and sketching) based pipeline of ExpressEdit. First, the user specifies their edit command in terms of natural language text and sketching on top of the frame. Then, the pipeline detects the temporal, spatial, and edit references from the text. It first interprets the temporal references to identify candidate edit segments. Then, for each candidate segment, it decides the exact spatial location based on the references in the text and the sketch. Lastly, it considers the references to edit operations and parameters to decide the applicable edit operations and their respective parameters.}
    \label{fig:pipeline_flow_example}
\end{teaserfigure}

\maketitle


\section{Introduction}

Informational videos introduce, explain, or demonstrate conceptual or procedural knowledge \cite{fiorella_what_2018, hove_characteristics_2014, truong_quickcut_2016}. They encompass a broad range of topics such as cooking, health, programming, and craft, and can be produced in various formats (e.g., lecture, tutorial, q\&a, demonstration, etc.) \cite{chi_democut_2013, kim_crowdsourcing_2014, guo_how_2014, betrancourt_why_2018}. Informational videos have become a popular source of knowledge for the general public due to their rich and engaging content \cite{kim_crowdsourcing_2014}, shared in various platforms such as YouTube \cite{youtube} and learning platforms such as Khan Academy \cite{khanacademy}, edX \cite{edx}, and Coursera \cite{coursera}.

However, producing informational videos involves a tedious and complex process that consists of multiple stages such as planning, scripting, filming, and editing \cite{chi_synthesis-assisted_2022, hua_lazycut_2005}. The editing stage is particularly tedious as it requires organizing footage, removing unnecessary parts, and finding and incorporating additional media assets \cite{chi_democut_2013} to ensure that the video delivers the intended knowledge in an informative and engaging manner.

While the popular commercial tools for video editing offer the necessary instruments to implement a variety of edits, for novices, these tools are difficult to learn and use, as they require great manual effort and have steep learning curves \cite{jokela_empirical_2007, chandler_cut_2004}. A stream of work has been introduced to make video editing less tedious and more efficient for novices such as automatically applying appropriate edits \cite{chi_democut_2013, berthouzoz_tools_2012, leake_computational_2017, ma_automated_2023, truong_quickcut_2016} or bootstrapping the editing process by generating videos \cite{chi_automatic_2021, chi_synthesis-assisted_2022, chi_automatic_2020, kalender_videolization_2018, zhong_helpviz_2021, udhayanan_recipe2video_2023, leake_generating_2020}. While these systems allow users to generate video edits efficiently, they do so in an automated or semi-automated manner which provides limited control over the process, which in turn inhibits the user's ability to express their editing intents.

On the other hand, natural language (NL) (e.g., text, speech) and other modalities (e.g., sketching, gestures) have been found to be effective for supporting intent expression and intuitive use for novices. Several multimodal interfaces were introduced for various creative tasks such as image editing \cite{laput_pixeltone_2013}, web styling \cite{kim_stylette_2022}, data visualization \cite{lee_multimodal_2018}, drawing \cite{van_der_kamp_gaze_2011}, and creative writing \cite{chung_talebrush_2022}. Although there has been work on multimodal interfaces for video production-related tasks like video review \cite{pavel_vidcrit_2016} or navigation \cite{gandhi_easy_2016, chang_rubyslippers_2021, zhao_rewind_2022, pavel_video_2014}, the design of multimodal interfaces for video editing has been under-explored.

We address this gap by investigating how multimodal interfaces can be leveraged in the informational video editing scenario. To explore the benefits and challenges of expressing video editing requests in a multimodal manner, we conducted a formative study with 10 video editors with diverse levels of expertise and collected 176 expressions of video editing requests in the form of NL texts, sketches, and media assets. We found that editors feel comfortable expressing their general editing requests through NL text (176 out of 176) and use sketching on top of the frame (78 out of 176) to indicate specific locations or regions of interest.

The results of the formative study motivated the design of \sysname{}, a multimodal interactive system for editing informational videos. It supports the expression of video editing requests through NL text and sketching on top of a frame, and is powered by a Computer Vision and Large Language Models-based technical pipeline that interprets and implements the edits by extracting three types of references from the multimodal edit command: (1) temporal location (e.g., \emph{``whenever a display appears''}), (2) spatial location within the frame (e.g., \emph{``towards the display''}), and (3) references to edit operations and their parameters (e.g., \emph{``slow zoom in''}) (Figure \ref{fig:pipeline_flow_example}). Additionally, \sysname{} provides a breakdown of the command into the aforementioned types of references, gives clear reasoning for each generated edit, and allows manual manipulation of those edits. 

We evaluated our technical pipeline by constructing a ground truth dataset from 50 selected expressions of video editing requests from the formative study\footnote{The dataset and our implementation of the pipeline can be found at \href{https://expressedit.kixlab.org/}{https://expressedit.kixlab.org/}}. The reference detection accuracy of our pipeline was higher than 0.68 for the three types of reference we support and yielded a 0.68 recall score on the temporal interpretation, 0.56 mIOU score on spatial interpretation, and 0.82 F-1 score on edit operation interpretation.

To evaluate the effectiveness of \sysname{} in editing informational videos, we conducted an observational study with 10 novice video editors. We found that \sysname{} facilitated the expression and implementation of video edits, provided a useful starting point to build upon, and allowed participants to feel more creative. Furthermore, the breakdown of the command and manual manipulation of the edits allowed users to iterate on their commands and polish their edits. 

Our paper makes the following contributions:
\begin{itemize}
    \item Formative study results demonstrating the role of NL text and sketching on top of the frame for expressing video editing requests. 
    \item Design of \sysname{}, a multimodal system that supports editing informational videos by enabling expression of video editing requests through NL text and sketching on top of the frame.
    \item A CV and LLM-based technical pipeline that understands and implements video edits by parsing and interpreting (1) temporal locations, (2) spatial locations, and (3) editing operations and their parameters from the NL text and sketch command.
    \item Results of the technical pipeline evaluation and the observational study that demonstrate the effectiveness and usefulness of \sysname{} in editing informational videos.
\end{itemize}

\section{Related Work}
Our work proposes a multimodal video editing system that allows users edit videos with natural language and sketching. We review related work on video editing systems and multimodal interfaces for creative tasks.

\subsection{Video Editing Systems}

Video editors have access to a plethora of editing tools \cite{adobepremiere, finalcutpro, imovie, descript, davinci} that support extensive set of functionalities that can be used to express and implement a variety of edits. Unfortunately, these tools require manual effort and have steep learning curves which make editing difficult for novices with limited knowledge \cite{hua_lazycut_2005, jahanlou_katika_2022, cabral_video_2017, casares_simplifying_2002, kiani_i_2020, jokela_empirical_2007, chandler_cut_2004}. 
Thus, various systems have been designed to help novices with video editing.
Earlier systems relied on metadata to simplify the editing process by providing multiple views with different semantic content and levels of abstraction \cite{casares_simplifying_2002}.
Several automatic approaches facilitated the video editing process by automatically applying edits based on predetermined markers \cite{chi_democut_2013}, placing transitions and cuts in interview videos \cite{berthouzoz_tools_2012}, adding visuals to audio travel podcasts \cite{xia_crosscast_2020}, selecting appropriate clips for dialogue-driven scenes \cite{leake_computational_2017}, adding lyric text to music videos \cite{ma_automated_2023}, and placing cuts by matching the user's voice-over annotations with relevant segments of the videos \cite{truong_quickcut_2016}.
Other systems bootstrapped the editing process by generating videos from documents \cite{chi_automatic_2021, chi_synthesis-assisted_2022}, web pages \cite{chi_automatic_2020, kalender_videolization_2018}, text-based instructions \cite{zhong_helpviz_2021}, recipe texts \cite{udhayanan_recipe2video_2023}, and articles \cite{leake_generating_2020} or synthesized talking head videos of puppets \cite{fried_puppet_2019} and used deep learning methods to automatically generate speech animations \cite{taylor_deep_2017}.
However, these automated approaches restrict the editor's control over the editing process by providing only predetermined input formats for interactions (e.g., markers, annotations), which in turn inhibits the expressiveness.

To alleviate the manual effort in video editing, researchers introduced text-based editing systems to allow users edit the video as if they were editing a text document \cite{wang_write--video_2019, fried_text-based_2019, yao_iterative_2020, huber_b-script_2019, pavel_sceneskim_2015, xia_crosscast_2020, berthouzoz_tools_2012, leake_computational_2017, huh_avscript_2023, fried_text-based_2019}. Such a method quickly found its application in popular video editing tools such as Adobe Premiere Pro \cite{adobepremiere}, Descript \cite{descript}, Imvidu \cite{imvidu}, and Remotion \cite{remotion}. Still, text-based editors mainly facilitate navigation and require manual effort to edit the videos end-to-end.
With \sysname{}'s multimodal interface, we aim to support novices by letting them easily express their video editing requests while providing sufficient control over the process and alleviating the manual effort required for editing.

\subsection{Multimodal Interaction}

Multimodal Interactions have been extensively researched by the HCI community and were found effective in reducing the burden and improving efficiency along with user satisfaction \cite{abioye_performance_2022, oviatt_when_2004, zimmerer_reducing_2022}. Related work has integrated multimodal interactions in various creative domains such as image editing \cite{laput_pixeltone_2013}, design \cite{pan_human-computer_2023}, creative writing \cite{chung_talebrush_2022}, drawing \cite{van_der_kamp_gaze_2011}, visualization \cite{su_natural_2021, lee_multimodal_2018}, and web styling \cite{kim_stylette_2022}. These approaches employ direct manipulation, speech-based, gesture-based, and gaze-based interactions to facilitate intent expression and intuitive use, as well as lowering the usability barrier for novices. Similarly, researchers have developed multimodal interactions for video-related tasks such as video review \cite{pavel_vidcrit_2016}, annotation \cite{ma_sketch-based_2012}, navigation \cite{gandhi_easy_2016, chang_rubyslippers_2021, zhao_rewind_2022, pavel_video_2014}, augmenting live storytelling, live presentations, and sports videos \cite{liao_realitytalk_2022, saquib_interactive_2019, chen_augmenting_2022}. In particular, there has been work to support video editing tasks such as annotating footage with speech and using the annotations to place cuts between footage \cite{truong_quickcut_2016}, mapping videos to 2D latent spaces to facilitate pattern identification \cite{lin_videomap_2022}, and utilizing pen gestures in editing process \cite{cabral_video_2017}.

However, these work fall short of addressing the video editing process holistically and provide limited insights into how to design multimodal interfaces for the combination of natural language (NL) and sketching interactions. Our work addresses this gap by conducting a formative study to learn about how video editors can utilize these modalities and presents a multimodal system \sysname{} that demonstrates the effectiveness of the approach.

On the other hand, the machine learning community has been active in researching NL-based video generation and editing \cite{qi_fatezero_2023, qin_instructvid2vid_2023, lee_soundini_2023, huang_style--video_2023, ho_imagen_2022, chai_stablevideo_2023, bar-tal_text2live_2022}. However, these work focus on the effectiveness of the AI model and usually support video editing tasks that manipulate the video on a pixel level (e.g., replacing objects, style transfer, etc). In this paper, we focus more on the interaction process with NL and sketching and address video editing tasks that frequently occur in informational videos.

\section{Formative Study}


To learn about (1) the role of natural language (NL) text and sketching in expressing video editing requests and (2) their use cases in editing informational videos, we conducted a formative study with video editors, where participants were asked to express their edit requests that would improve the informativeness and engagement of a given video. We call these expressions \textit{edit commands}.


\subsection{Participants}
We recruited 10 video editors (3 females, 7 males, mean age $24.7$) with prior experience editing informational videos. We recruited 5 novices who had edited at least two videos and watched informational videos regularly and 5 experienced editors who had edited at least 20 videos and 5 informational videos (Table \ref{tab:fs_participants}). We recruited both novice and experienced editors to cover a diverse range of edit commands since edit expressions such as attention to detail and vocabulary used can vary depending on the participant's editing expertise and ensured that novice editors are also regular viewers of informational videos so that they have an understanding of the types of edits that should be included in such videos. In the recruitment form, we asked about participants' editing experience, the topics of informational videos they have edited or watched, and their demographic information. The editors were recruited through Upwork \cite{upwork}, a freelancing platform, and university community postings. Novice and experienced participants were compensated with 30,000 KRW (approximately 25 USD) and 50,000 KRW (approximately 40 USD) for a 100-minute study, respectively.

\subsection{Study Materials}
We chose five archived informational live streams as raw footage for the study. The videos covered various topics, knowledge characteristics (i.e., procedural or conceptual), and the main channel of information (i.e., visual or verbal) (Table \ref{tab:fs_videos}). We chose archived live streams as they are usually unedited and closely resemble a continuous stream of raw footage, which allows for tasks closer to real-world video editing settings. We assigned the videos to each participant that best fit their interests as closely as possible (Table \ref{tab:fs_participants}). 

To allow participants to express their edit commands in both text and sketch, we used Google Slides \cite{googleslides}, a popular slide authoring tool. We chose the tool because of its functionalities of adding text, images, and shapes, which could be used in expressing edit commands. Participants were also allowed to take a screenshot of a frame of the video and sketch over it. 

\subsection{Procedure}

The study was conducted either online through Zoom \cite{zoom} or in person depending on the participant's preference (Table \ref{tab:fs_participants}). During the study, the participants were first asked to skim through the given video and its transcript for 30 minutes to become familiar with the content. Then, they were given a quick tutorial on the basic functionalities of Google Slides such as adding text and shapes. For the next hour, participants were tasked to produce 20 semantically unique expressions of edit commands as if they were explaining them to another video editor. We set 20 commands as a target number but did not prolong the study if participants could not reach the amount. They were free to switch between the video and the slides during the study and were asked to put each edit command into a single slide. After participants completed the main task, we conducted a semi-structured interview for about 10 minutes to learn about the participant's experience performing the given task compared to their previous experiences.

\subsection{Findings and Analysis}
We analyzed the collected edit expressions in terms of their patterns and usages. 
Below, we summarize the findings from the study.

\subsubsection{Expressing edit commands with multi-modalities}

Overall, the participants expressed edit commands using various modalities: NL text, sketch, image, and graphics. All of the commands contained NL text to express the edit they want to implement, such as \textit{``18:08 - cut at the point where he starts talking about leeks''}. The participants also added sketches on top of video frames to refer to part of the frame, such as by adding a rectangular shape tool with an accompanying text \textit{``Reduce the prominence of white in \textbf{these parts} of the video due to overexposure.''} They also added sketches, images, and graphics to describe the content they wanted to add, such as images of an okay-sign icon with the accompanied text \textit{``Add in \textbf{hand emoji} for delicious in the top left corner whenever he says the word ``delicious'' for a split second''}.

Most participants felt comfortable expressing their edit commands with the given tool. For P5 and P9 it was a common practice to either plan out the edits with NL and sketching or communicate the edit descriptions through NL in collaborative settings. 
However, P10 and P7 noted that it would be cumbersome to articulate all the details of the edit with just a text. P9 also mentioned that he would prefer to work on a tablet for sketching, and P5 wished to use more sophisticated image editing tools (e.g., Adobe Photoshop \cite{adobephotoshop}) to express the exact visual effects he wanted.



\subsubsection{Participants consistently referenced moments in the video with NL text.}

Most frequently, participants specified one or several timestamps in the video where they wanted to apply the edit, resembling how they would apply edits on the timeline in existing video editing systems. However, when participants did not know the exact moments or wanted to refer to multiple moments with the same edit, they opted to give higher-level references to the visual content of the video (e.g., actions, objects, or their descriptions) or verbal content of the video (e.g., transcript or sounds).

\subsubsection{Participants used both NL text and sketching on top of the frame to reference the spatial location of the edits.}

The participants specified a spatial location in NL text (e.g., ``top-left corner'', ``zoom into pan'') or sketches on top of the relevant frames. They mainly used visuals to directly illustrate the edit, similar to how they would implement edits on canvas in typical video editing systems. Among 176 multimodal edit commands collected, 97 contained visuals such as sketches, images, and graphics. Out of the 97 visual commands, 78 contained a frame of the video as a reference and a sketch on top of it (e.g., free-form, shapes, images).


\subsubsection{Participants used NL text to refer to edit operations and their parameters.}

There were several types of references to edit operations and their parameters. Most commonly, the participants referred to them by directly including the name of the operation (e.g., cut, text, image, etc.) that is supported in existing video editing systems. Alternatively, they mentioned the main purpose or intended effect of the edit (e.g., highlight, emphasize, focus). In terms of the parameters of the edit operations, they mostly gave general descriptions (e.g., large text, slow zoom) of the intended effects. Occasionally, experienced editors mentioned precise numbers or specific terminology (e.g., ``slow down by 50\%'', ``jump cut''). For edits that contained additional media (e.g., text content, image source), participants used references to the transcript or explicitly described the content.

\subsubsection{Participants frequently iterated on their edit commands to make them clearer.}

During the study, we observed that participants frequently revisited and refined their edit commands, for example, to make them more precise or to keep the consistency between multiple commands. However, they mentioned that the task setting lacked the real-time iterative nature of video editing, where they could repeatedly improve the edits they had implemented. 

\subsection {Design Goals}
Based on the findings from the formative study, we identified three design goals for a system that supports video editing via natural language (NL) text and sketching:

\begin{itemize}
    \item DG-1: Enable expression of edit commands through natural language and sketching on top of the frame;
    \item DG-2: Based on the multimodal edit command, support interpretation of temporal locations, spatial locations within the frame, and editing operations with parameters;
    \item DG-3: Support iteration on multimodal edit commands and manual editing of interpretation results;
\end{itemize}

\setlength{\belowcaptionskip}{15pt}

\begin{table}[t]
    \centering
    \resizebox{\columnwidth}{!}{
        \begin{tabular}{c|c|c|c|c} \hline 
            Participant &
            Experience &
            Edited inform. videos \# &
            Assigned video & 
            Study format \\ \hline 
            P1& Novice &        0    & FV1 & In Person  \\ \hline 
            P2& Novice &        0    & FV2 & In Person  \\ \hline 
            P3& Novice &        0    & FV3 & In Person  \\ \hline 
            P4& Novice &        0     & FV4 & Online    \\ \hline 
            P5& Novice &        0     & FV5 & Online    \\ \hline 
            P6& Experienced &   12   & FV1 & Online    \\ \hline 
            P7& Experienced &   13   & FV2 & In Person  \\ \hline 
            P8& Experienced &   10   & FV3 & In Person  \\ \hline 
            P9& Experienced &   120 & FV4 & In Person  \\ \hline 
            P10& Experienced &  13   & FV5 & In Person  \\ \hline
        \end{tabular}
    }
    \caption{Information about formative study participants including their experience level, the number of informational videos they edited, the assigned live stream video for the formative study, and the format of the study.}

    \Description{The table shows the information (i.e., experience level, number of information videos edited, the assigned formative study video, and format of the study (online or in person) ) about formative study participants. There are 10 participants in total where five are novices who have never edited informational videos but at least have edited two videos and five are experienced editors who edited at least ten informational videos. The selected formative study videos were assigned randomly and three of the participants (2 novices and 1 experienced editor) participated in the study remotely, while others attended the study in person.}
    \label{tab:fs_participants}
\end{table}

\begin{table*} [ht!]
    \centering
    \resizebox{.9\textwidth}{!}{%
        \begin{tabular}{c|c|c|c|c} \hline 
             Video & Topics & Knowledge Characteristics & Content Format & Title \\ \hline 
             FV1&   cooking&procedural&  visual \& verbal &At Home for the Holidays with Gordon Ramsay \cite{ramsey_1} \\ \hline 
             FV2&   educational&procedural&  verbal &Language Learning Live Stream \cite{kaufmann_1} \\ \hline 
             FV3&   programming&procedural&  visual \& verbal &Livestream: Getting Started with C++ (Episode 1) \cite{codeacademy_1} \\ \hline 
             FV4&   health&conceptual&  verbal& Surgeon does Live QA | Hair Loss Awareness Month \cite{linkov_1} \\ \hline 
             FV5&   product review&conceptual&  visual \& verbal &Microsoft Surface Go - Classic LIVE Unboxing \cite{linus_1} \\ \hline
        \end{tabular}
    }
    \caption{
        The table shows the information for live-stream videos selected for the formative study including topics, knowledge characteristics, content formats, and titles with links.
    }
    \Description{The table presents the descriptions of the live-stream videos used in the formative study. There are five videos from diverse informational topics (e.g., cooking, educational, programming, health, and product review). Additionally, the selected videos cover different combinations of knowledge characteristics (procedural or conceptual) and content formats (visual or verbal). It mentions the titles of the videos in the last column.}
    \label{tab:fs_videos}
\end{table*}

\section{\sysname{}: Interface}

\begin{figure*}
  \includegraphics[width=\textwidth]{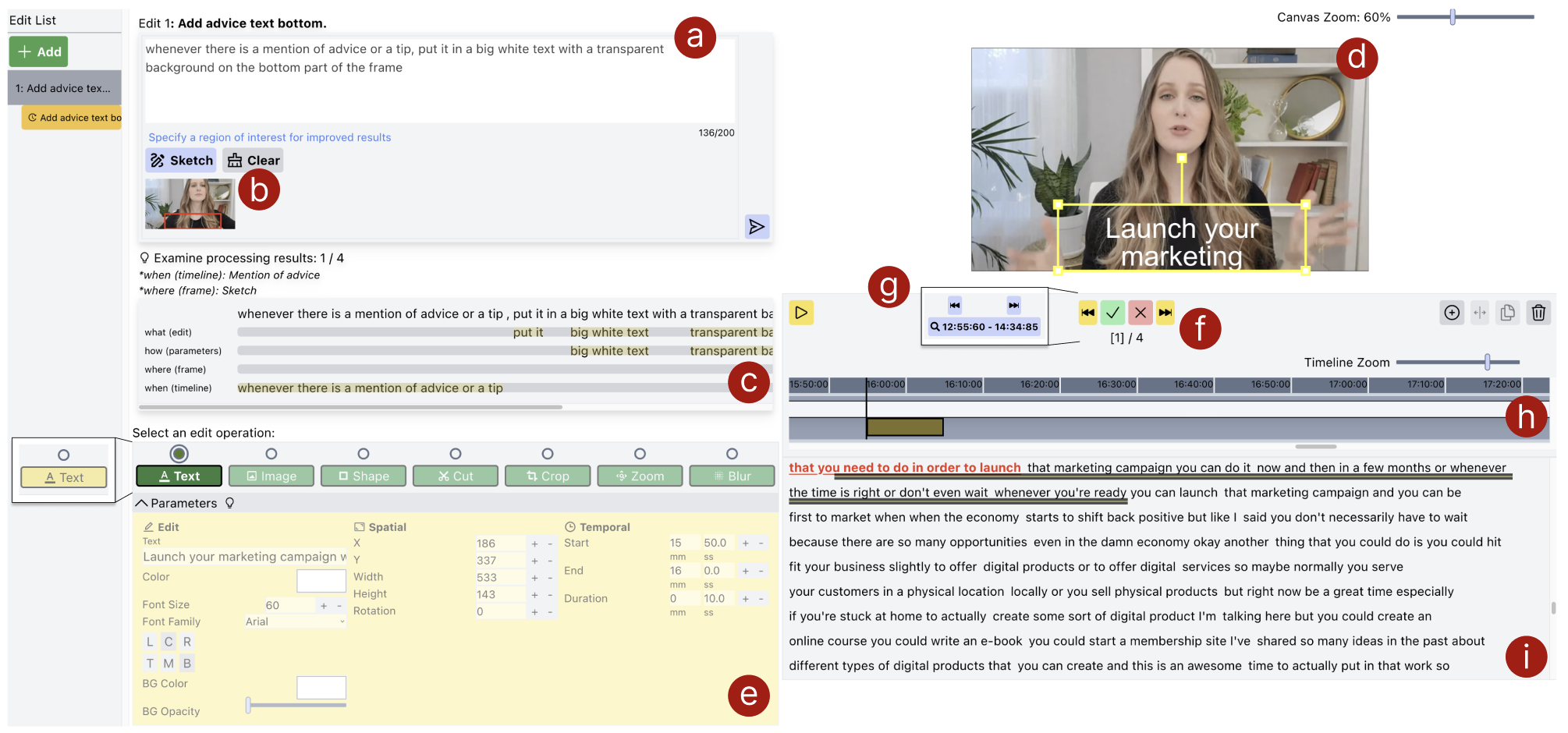}
  \caption{With \sysname{}, the user can (a) input an edit command using natural language in the \textit{edit description box} and (b) optionally specify the location using the \textit{sketch} function. The system analyzes the request and (c) shows the parts of the NL prompt that correspond to the user's intended edit operation, parameters, spatial location, and temporal location. (d) The \textit{editor canvas} shows the preview of the edits and allows clicking and dragging. (e) Users can also manually adjust the resulting edit operation as well as its given parameters. (f) Users can navigate through the edit suggestions and accept or decline, (g) as well as quickly navigate through the video timeline. (h, i) The \textit{timeline} and \textit{transcript} shows the temporal location of applied edits and edit suggestions.
  }
  \Description{An interface of \sysname{}. It contains a tab-like column on the left that stores users' commands. On the top-left users can put edit expressions and see the processing results in the timeline on the right side. The interface also contains an edit panel on the left bottom where users can choose an edit or change the specific parameters of the edit. On the top-right, there is a canvas that users can use to specify spatial locations of the edits such as texts or images. Lastly, on the bottom right, there is a transcript of the video where users can navigate the video and see all the edits.}
  \label{fig:system_scenario}
\end{figure*}

Following the identified design goals, we designed \sysname{}, a multimodal video editing system that allows the expression of edit commands through natural language and sketching on top of the frame (NL\&S). Our system interprets the user's request in the form of NL\&S (DG-1) and suggests a set of edits with temporal location (when in the video), spatial location (where in the video frame), and edit operation \& parameters (which edit and how) (DG-2). To better understand the suggested edits and iterate on the NL\&S command, users can examine the summary of the system's processing results that shows how each part of the NL part of the command was interpreted by the system and the reasoning behind each suggested edit's temporal and spatial location (DG-3). Additionally, users can manually adjust the suggested edits by the system or create their own edits (DG-3).

\subsection{User Scenario}


To illustrate how \sysname{} can be used, let's follow Lia, a businesswoman and a YouTube creator who wants to edit her video about entrepreneurship. She recorded a talking-head video (i.e., a popular style for informational videos that centers on the speaker's face and upper body \cite{fried_text-based_2019}) that she wants to make more informative and engaging using \sysname{}.

\subsubsection{Creating a new edit}

To start editing the video, Lia first uploads her recorded footage with the transcript to \sysname{} and comes up with the first edit that she wants to implement. She presses the \textbf{Add} button on the \textbf{Edit list} panel and creates a new layer on top of the video where she can apply her edits. Edits within a single layer will be of a single edit operation and cannot intersect with each other temporally.

\subsubsection{Describing the edit with NL \& Sketch}

Lia decides to first add text captions whenever she mentions valuable advice or tips. In the \textbf{Edit description} (Figure~\ref{fig:system_scenario}a), she types \emph{``whenever there is a mention of advice or a tip, put it in a big white text with a transparent background on the bottom part of the frame''}. Additionally, she specifies the exact part of the frame where she wants the text to appear using the \textbf{Sketch} function (Figure~\ref{fig:system_scenario}b) and draws the bounding box on the bottom half of the frame. She presses \textbf{Enter} to process the NL\&S request and gets (1) a breakdown of her command on the \textbf{Examine} panel (Figure~\ref{fig:system_scenario}c), (2) a suggested edit operation highlighted by yellow on the \textbf{Edit Operation} panel, (3) set of edits on synchronized \textbf{Timeline} (Figure~\ref{fig:system_scenario}h) and \textbf{Transcript} (Figure~\ref{fig:system_scenario}i) that indicate where the edits should be applied. Additionally, the system summarizes Lia's description and adds a new entry under the current edit in \textbf{Edit List} panel that saves her description for future revisits \& refinements. This happens every time Lia makes a new description for the edit and the suggestions (e.g., edit operations, edits) will be saved in the corresponding entry. The system automatically chooses the suggested edit operation and matches the player's position to the start of the first suggested edit.

\subsubsection{Examining the results}

To determine if the system interpreted her NL\&S command correctly, Lia examines the breakdown of the NL part of her command, along with the elements of the interface that are highlighted with yellow color. She sees that on the first row of the breakdown \textbf{what (edit)} the \emph{"text"} is outlined and notices that the \textit{Text} operation was automatically chosen in the \textbf {Edit operation} panel. In the second row \textbf{how (parameters)}, the part of the NL command with a description of the appearance of the text is highlighted \emph {``a big white text with a transparent background''}. She also confirms that the edit parameters are set appropriately by looking at the \textbf{Editor Canvas} (Figure~\ref{fig:system_scenario}d) and \textbf{Parameters Panel} (Figure~\ref{fig:system_scenario}e). The part of the NL command referencing the edit's spatial location within the frame is highlighted on the third row \textbf{where (frame)}. The outlined text \emph{``the bottom half of the frame''} and the position of the edit on the \textbf{Editor Canvas} assures her that the location was interpreted correctly. She also reads the quick reasoning for the spatial location on the top of the breakdown rows which says \emph{Sketch} indicating that her sketch was used to decide the location of her edit. Lastly, Lia sees that the reference \emph{"whenever there is a mention of advice or a tip"} is outlined on the fourth row labeled \textbf{when (timeline)} and glances at the snippet of the transcript of the video in the \textbf{Transcript} panel that highlighted the exact moment where she was talking about "marketing campaign". She also reads the quick reasoning for suggesting the edit on the top of the breakdown rows: \emph{"Mention of advice"}. Thus, she ensured that the NL\&S command was interpreted as she expected.

\subsubsection {Manual Manipulation \& Editing}

To go through all the suggested edits in the video, Lia jumps between them with \textbf{Previous} and \textbf{Next} buttons (Figure~\ref{fig:system_scenario}g). She glances at the transcript for each edit and judges whether to insert a text on the video in that segment or not and expresses her decision with \textbf{Accept} and \textbf{Reject} buttons (Figure~\ref{fig:system_scenario}f). After making all the decisions, she is left with four text edits that are applied to the video. She remembers that there was one more important piece of advice missed by the initial set of suggested edits, so she roughly navigates to the part of the video where she thinks the segment is in the \textbf{Timeline} and presses the \textbf{Search more} button (Figure~\ref{fig:system_scenario}g) that gives her the exact moment when she mentions the advice. She accepts the suggested edit and goes on to manually adjust its parameters. She specifies the exact temporal boundaries of the edit by dragging them on the \textbf{Transcript}, then adjusts the position and size of the text on the \textbf{Editor Canvas}, and slightly rephrases the text content in the \textbf{Parameters} panel. After finalizing the edit, she moves on to the next edit that she has by pressing the \textbf{Next} button that now jumps between the applied edits.

\subsection{Edit operations}\label{sec:edit_operations}

\sysname{} supports 7 edit operations: (1) text insertion, (2) image insertion, (3) shape insertion (circle, rectangle, star), (4) cutting out segments of the video, (5) zooming in/out, (6) cropping the video, and (7) blurring the video. These were the 7 visual edit operations that formative study participants frequently mentioned in their edit commands. We decided to focus only on visual edit operations as they cover the important types of parameters (temporal, spatial, edit-specific) that commonly appear in other kinds of edits (e.g. audio-related edits, coloring edits). We believe that this set of edit operations effectively demonstrates the feasibility of implementing various kinds of edits based on NL\&S commands. 

\subsection {Implementation}

\sysname{} is implemented as a Web-based React \cite{react} application powered by KonvaJS \cite{konvajs} for canvas manipulations and MobX \cite{mobx} for state management. The backend server was implemented using Flask \cite{flask}, which hosted the videos along with their transcript and processed users' requests. All the videos and transcripts that were used in this work were obtained from YouTube with youtube-dl package \cite{youtubedl} for Python. 

\section{\sysname{}: Pipeline}

Based on the findings from our formative study, we designed our pipeline to interpret natural language edit commands. We also support sketching on top of the frame to allow users more effectively convey the spatial location for a particular edit along with their natural language command. We also performed technical evaluation of the pipeline by constructing a ground truth dataset from 50 multimodal edit commands selected from the formative study.

\subsection{Video Pre-processing}
\begin{figure*}[ht!] 
    \centering
    \includegraphics[width=.9\textwidth]{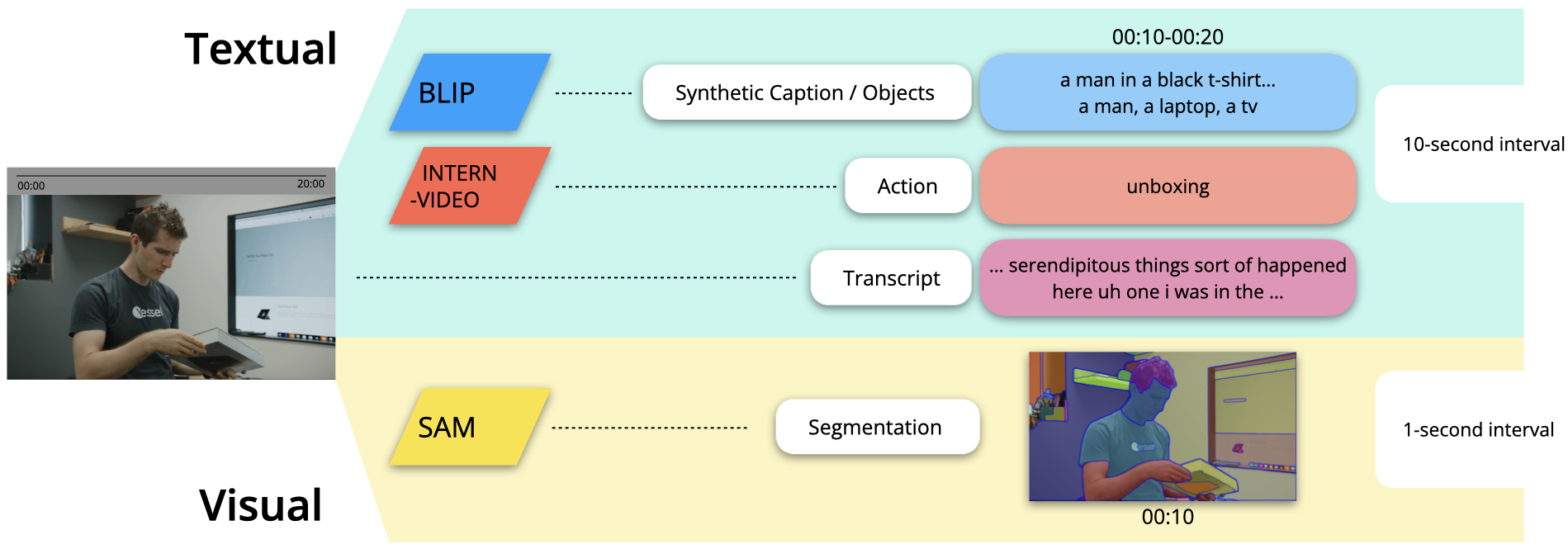}
    \caption{The offline component of the pipeline pre-processes the video and extracts textual and visual metadata.}
    \Description{The figure shows the walkthrough of the pre-processing stage of the pipeline in detail. First, it divides the video into 10-second segments and extracts textual descriptions. Additionally, it extracts segmentations for frames every 1.0 seconds.}
    \label{fig:pipeline-pre}
\end{figure*}

In order to facilitate the real-time system interactions, we perform pre-processing to extract frame-level and clip-level metadata from a video that \sysname{} uses to reason about video context during the interaction (Figure~\ref{fig:pipeline-pre}). For the frame-level data, we sample video frame every 1.0 seconds and use the Segment Anything model \cite{kirillov_segment_2023} to run automatic instance segmentation at two levels of granularity. We then threshold the segmentations based on instance crop size to remove instance masks that are too small. For the clip-level metadata, we run activity recognition on 10-second clips using InternVideo \cite{wang_internvideo_2022}. We also employ an image captioning model BLIP-2 \cite{li_blip-2_2023} to generate textual descriptions of the video contents and list the most salient objects in the scene at each second in the 10-second intervals. We then use GPT-3.5 \cite{brown_language_2020} to summarize the resulting captions across the 10-second clips to remove redundant information about the visual content. 

\subsection{Parsing Edit Command}

 We first use GPT-4 \cite{openai_gpt-4_2023} to parse the NL command and divide it into the following types of references as illustrated in Figure~\ref{fig:pipeline-post}: 
\begin{enumerate}
    \item \textbf{Temporal reference}: any information in the NL command that could refer to a segment of the video;
    \item \textbf{Spatial reference}: any information in the NL command that could refer to location or region in the video frame;
    \item \textbf{Edit Operation reference}: any information in the NL command that could indicate an edit operation to use;
    \item \textbf{Edit Parameter reference}: any information in the NL command that could refer to specific parameters of edit operation that was determined;
\end{enumerate}
We then use GPT-4 to classify the edit operation that is most suitable for the request based on our system's available operations: ``text'', ``image'', ``shape'', ``blur'', ``cut'', ``crop'', or ``zoom.'' These references are then further interpreted and linked to sections within the video as outlined in the following sections.

\subsection{Temporal Interpretation}
In the next stage of the pipeline, the temporal references are decomposed into the following categories (Figure~\ref{fig:pipeline-post}):
\begin{enumerate}
    \item  \textit{Positional} references can be either distinct timecodes or abstract temporal designations (e.g., ``intro'', ``ending'');
    \item \textit{Transcript-based} references constitute either direct or indirect references to the video's transcript;
    \item \textit{Video-based} labels are assigned to descriptions that refer to actions or visual descriptions specific to the video;
\end{enumerate} Given the volume of video metadata, if the label is either \textit{transcript-based} or \textit{video-based} we filter the 10 most relevant temporal segments of the textual clip-level metadata using cosine similarity along the dense captions and transcript segments, respectively using SentenceTransformers framework \cite{sentencetransformers}. We then use GPT-4 to compile all segments of the video that match the temporal reference and pass the candidate segments further along the pipeline.

\subsection{Spatial Interpretation}

Upon obtaining a list of candidate temporal segments, the pipeline initiates the parsing of spatial references. As shown in Figure~\ref{fig:pipeline-post}, we categorize these references as:
\begin{enumerate}
    \item \textit{Visual-Content Dependent} references to specific to objects, elements, or regions within the video frame;
    \item \textit{Visual-Content Independent} references to specific locations or positions that are relative to the frame, but not contingent on the video's visual content.
\end{enumerate}

For segments with ``visual-content dependent'' information, we extract representative frames from the center of the time-coded range. We then obtain the corresponding instance crops from the frame-level metadata for each frame. If the user provides a sketch, the text and sketch are encoded into a shared embedding space alongside all instance crops using CLIP \cite{radford_learning_2021}. Without a user-provided sketch, only the textual reference and instance crops are encoded in the embedding space. The spatial location within the frame that aligns most closely with the parsed command and/or sketch is determined by finding the instance crop having the highest cosine similarity to the user's input.

In the absence of ``visual-content dependent'' references, if a sketch is provided, it becomes the candidate spatial location. If neither is available, the full frame is returned as the final spatial location. Additionally, if the NL command contains only ``visual-content independent'' spatial references, we employ GPT-4 to refine and resize the region of interest based on the command and frame boundaries.

\subsection{Edit Operation and Parameters Interpretation}

Lastly, we leverage the relevant segments from the NL command to modify the parameters corresponding to each predicted edit operation. These modification requests can be grouped into three categories:
\begin{itemize} 
\item \textit{Explicit} specifications for parameters like ``12px'' or ``Introduction''; 
\item \textit{Relative} adjustments relative to current settings such as ``5 seconds longer'' or ``10\% less''; 
\item \textit{Abstract} (i.e., general) directives that don't have specific values associated to them, like ``shorter'' or ``longer''; 
\end{itemize}

Particular emphasis is given to operations involving text and images. For these operations, we provide the command, its context, and the relevant video content to guide the generation process. In the case of text, this aids in determining the textual display within the video. For images, it assists in formulating an appropriate search query to source the required images for the video.

\subsection{Implementation}

\begin{figure*}[ht!] 
    \centering
    \includegraphics[width=.9\textwidth]{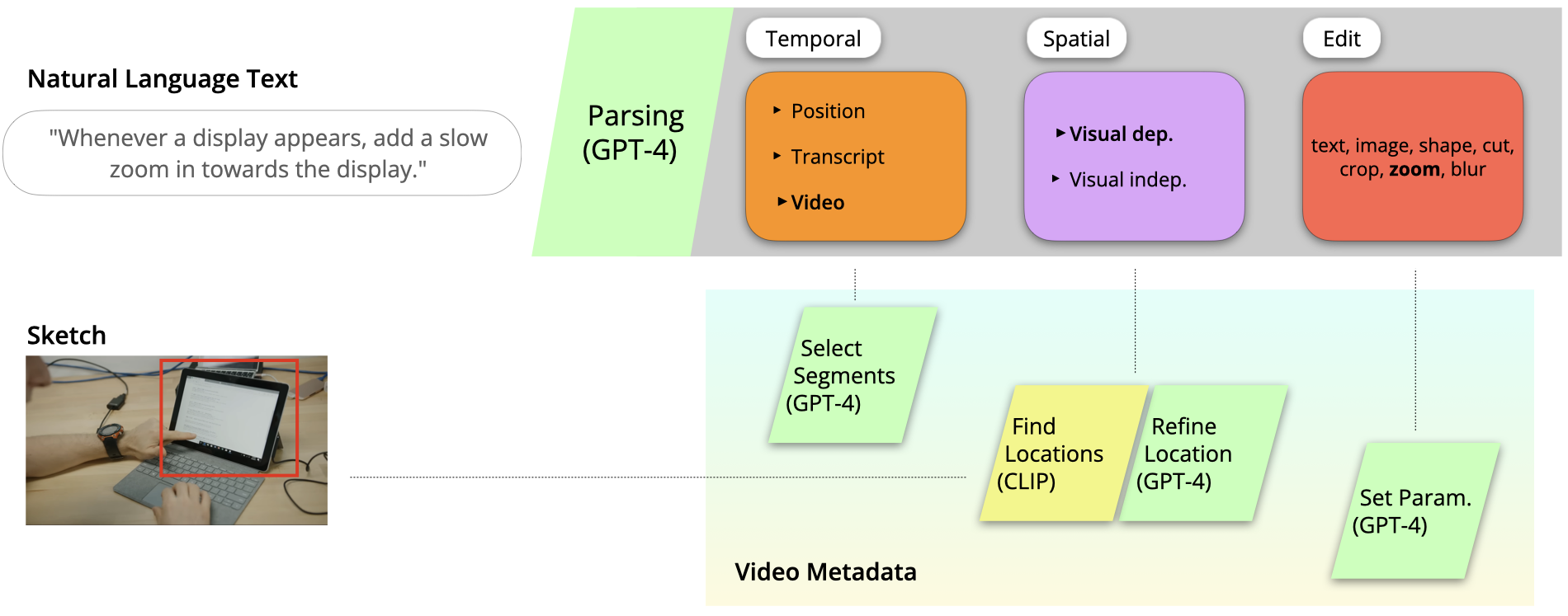}
    \caption{The online component of the pipeline uses GPT-4 and CLIP to interpret NL text and sketching edit command.}
    \Description{The figure shows the walkthrough of the online component of the pipeline. It receives the edit command and based on the pre-processed data, uses GPT-4, to first identify temporal, spatial, and edit references, then find the applicable edit operation. Further, it interprets the temporal references into candidate edit segments, and then identifies spatial locations for each. Lastly, it interprets the edit parameters references to construct each suggested edit.}
    \label{fig:pipeline-post}
\end{figure*}

Our implementation of the pipeline can be divided into an offline pre-processing component (Figure \ref{fig:pipeline-pre}) and an online component (Figure \ref{fig:pipeline-post}). While the offline component is run locally with all the necessary models, the online component was implemented as a Flask server where we used LangChain framework \cite{langchain} to perform edit command processing. The prompts used for each stage of the pipeline can be found in Appendix~\ref{apx:prompts}.


\subsection{Pipeline Evaluation}

We evaluated the performance of our CV and LLM-based pipeline through comparison with the ground truth dataset that consisted of 50 multimodal edit commands selected from the formative study. Table~\ref{tab:pipeline_results} shows the summary of the results for parsing accuracy of the NL commands into (1) temporal, (2) spatial, (3) operational, and (4) parametric parts; and interpretation accuracy for each type of reference. For all the experiments we used \textit{gpt-4-0613} with temperature 0.0.

\subsubsection{Ground Truth construction}
To construct our ground truth dataset, we selected 50 expressions of editing requests (out of 179) from our formative study dataset. Our inclusion criteria were as follows:
\begin{enumerate}
    \item The edit request's operation reference maps to one of the system's supported editing operations.
    \item The edit request's temporal reference is either \textit{positional}, \textit{transcript-based}, or \textit{video-based}.
    \item The edit request is fully self-contained. In other words, it does not reference other edit commands. 
\end{enumerate}

Two authors analyzed each multimodal edit command to extract parts that refer to temporal locations, spatial locations, edit operations, and parameters from the textual component. Then, they interpreted each reference by (1) detecting moments in the video, (2) choosing appropriate spatial locations within the frame, and (3) selecting suitable edit operations. The full analysis and interpretation procedure is listed in Appendix~\ref{apx:gt}

\subsubsection{Metrics}
To measure the performance of the parsing and interpretation stages of the pipeline we calculate cosine similarities, F-1 scores (along with precision and recall), and mean Intersection-Over-Union (mIOU) between predicted results and ground truth. 

For each type of reference (i.e., temporal, spatial, operational, parameter) we compute the text embeddings of both the predicted parsing result and the ground truth and calculate the cosine similarity between them with SentenceTransformers framework \cite{sentencetransformers}. 

To assess the accuracy of the interpretation of the temporal references, we compute F-1 scores between predicted and ground-truth segments similar to Yang et al. \cite{yang_catchlive_2022} Since our pipeline mainly predicts short segments of the video with a 10-second duration and the small time differences between the predicted segment and ground truth are insignificant (as users usually refine the boundaries anyway), we count the predicted segment as true positive if it intersects with the ground truth segment. Generally, it is also useful if the predicted segment is near the desired moment in the video, thus we also report the F-1 scores with relaxed true positive criteria: the distance between the predicted segment and ground truth is within 10 seconds.

As spatial interpretation happens for each predicted segment, we isolated the performance of the spatial interpretation by computing the mIOU between the predicted and ground truth locations for the ground truth segments of each data point. Since the ground truth bounding boxes of locations are approximate and the exact width and height of the boxes usually depend on the content, we also report the ratio of mIOU scores larger than 0.1 which ensures that there is some intersection between the prediction and ground truth.

Since our ground truth dataset contains edit commands that required a combination of edit operations, our pipeline is designed to suggest multiple of them. Thus, we calculate F-1 scores to measure the edit operation interpretation performance. Since all the edit operations are from the set of seven operations that \sysname{} supports (Section~\ref{sec:edit_operations}), we consider the prediction to be true positive if it exactly matches one of the operations in the ground truth.

\subsubsection{Results}

Our pipeline achieved reasonable average cosine similarity scores for all the reference types that the system supports. The parsing spatial reference parts from the NL command have the lowest score and highest variability among all (M=0.68, STD=0.30), which might be due to the fact that we had fewer data points with NL spatial references (22 out of 50) and ground truth contained generic references such as ``over the video'' and ``on the screen'' that did not contribute concrete specifications to the spatial location. The scores for parsing temporal references (M=0.79, STD=0.27), edit operation references (M=0.72, STD=0.25), and parameter references (M=0.74, STD=0.20) have relatively similar standard deviations with operation references having the lower score. 
The average F-1 score for interpreting edit operation references is 0.82 (STD=0.32) and the precision and recall are 0.84 (STD=0.32) and 0.83 (STD=0.33), respectively. For interpreting temporal references, the recall is 0.68 (STD=0.43) and 0.71 (STD=0.41) with a 10-second margin. Although the F-1 score (M=0.55, STD=0.40) and precision (M=0.52, STD=0.40) are much lower, they are less important compared to recall as finding and making sure that all the important moments in the video are covered (recall) is usually much more challenging and time-consuming compared to evaluating if a short snippet is relevant (precision). Thus, even with lower precision but reasonable recall, our pipeline can still be useful for video editing tasks \sysname{} supports. 
Finally, the average mIOU for interpreting spatial references is 0.56 (STD=0.40) and the ratio of mIOUs larger than 0.1 is 0.86, which shows that the pipeline is generating reasonable bounding boxes for the edits.



\begin{table*}[ht!]
\centering
\resizebox{0.8\textwidth}{!}{%
\begin{tabular}{l|c|cccc}
    \hline
    \multirow{2}{*}{Reference Type} &
      \multirow{2}{*}{\begin{tabular}[c]{@{}c@{}}Parsing Performance\\ Average (STD)\end{tabular}} &
      \multicolumn{4}{c}{\begin{tabular}[c]{@{}c@{}}Interpretation Performance\\ Average (STD)\end{tabular}} \\ \cline{3-6} 
     &
       &
      \multicolumn{1}{l|}{F1} &
      \multicolumn{1}{l|}{Precision} &
      \multicolumn{1}{l|}{Recall} &
      mIOU / mIOU (T=0.1) \\ \hline
    
    Temporal (when in the video) &
      0.79 (0.27) &
      \multicolumn{1}{l|}{0.55 (0.40)} &
      \multicolumn{1}{l|}{0.52 (0.40)} &
      \multicolumn{1}{l|}{0.68 (0.43)} &
      - \\ \hline
    \begin{tabular}[c]{@{}l@{}}Temporal (when in the video)\\ \textit{margin=10s}\end{tabular} &
      - &
      \multicolumn{1}{l|}{0.57 (0.38)} &
      \multicolumn{1}{l|}{0.56 (0.39)} &
      \multicolumn{1}{l|}{0.71 (0.41)} &
      - \\ \hline
      
    Spatial (where in the frame) &
      0.68 (0.30) &
      \multicolumn{1}{l|}{-} &
      \multicolumn{1}{l|}{-} &
      \multicolumn{1}{l|}{-} &
      0.56 (0.40) / 0.86 \\ \hline
    Edit Operation &
      0.72 (0.25) &
      \multicolumn{1}{l|}{0.82 (0.32)} &
      \multicolumn{1}{l|}{0.84 (0.32)} &
      \multicolumn{1}{l|}{0.83 (0.33)} &
      - \\ \hline
    Edit Parameters &
      0.74 (0.20) &
      \multicolumn{1}{l|}{-} &
      \multicolumn{1}{l|}{-} &
      \multicolumn{1}{l|}{-} &
      - \\ \hline
\end{tabular}
}
\caption{Performance of the CV\&LLM-based technical pipeline in terms of parsing and interpretation of (1) temporal, (2) spatial, (3) edit operation, and (4) edit parameters. We calculated the spatial interpretation performance separately with ground truth edit segments instead of predicted segments.
}
\Description{The table shows the performance of CV\&LLM-based technical pipeline for interpreting multimodal edit commands. The average parsing performance across all types of references is above 0.68. For performance of interpretations is calculated in terms of F1-score, precision, and recall. Most importantly, the recall for temporal references is 0.68, which means that the pipeline finds most of the requested moments in the video well. The performance of the pipeline on the spatial references is 0.56 in terms of mean Intersection-Over-Union, however, the predicted locations were intersecting at 0.86 percent on average. Lastly, the edit operation interpretation accuracy is above 0.82 across all F1 scores, precision, and recall.}
\label{tab:pipeline_results}
\end{table*}

\section{User Evaluation}

\sysname{}'s main features are designed to enhance the expression of edit commands and facilitate their implementation. We conducted an observational study with novice video editors, who are more likely to face challenges in expressing and implementing their editing ideas. In particular, we aim to address the following research questions:

\begin{itemize}
    \item RQ-1: How do users edit videos with \sysname{}?
    \item RQ-2: How well does \sysname{} understand and implement multimodal commands?
    \item RQ-3: How useful is \sysname{} for novices in editing informational videos?

\end{itemize}

\subsection{Participants}
We recruited 10 participants (5 females, 5 males, mean age $20.4$) who identified themselves as novice editors through university community service postings. The reported number of years of experience and the number of videos each participant edited prior to the study are shown in Table \ref{tab:eval_participants}. We also verified that the participants are regular viewers of informational videos and can write and speak in English fluently, as our system accepts NL commands in English. 

For the study, we provided two video footage and reference video pairs, each with similar topics and characteristics. Participants were tasked with making the given video footage more engaging and informative using their own ideas or ideas derived from the provided reference video. Similar to videos for the formative study, the selected footage videos varied in terms of (1) knowledge characteristics (procedural/conceptual), and (2) formats (visual/verbal), as outlined in Table \ref{tab:eval_videos}. The footage video contained a minimal number of edits, while the reference video contained a range of edit ideas relevant to the footage and supported by our system.

\subsection{Procedure}

The study was conducted in person and the entire session (audio and screen) was recorded. After the introduction of the study, the participants watched a 10-minute-long tutorial on the system and had 10 more minutes to complete a set of simple tasks where they were asked to use each edit operation we support at least once and make at least two NL\&S edit commands to the system. After the tutorial, participants skimmed through the assigned footage and the reference for 10 minutes and performed the main task of improving the video by making it more engaging and informative for 40 minutes. They were encouraged to think aloud and describe the intentions behind their actions as they were implementing edits. Right after the main task, participants completed the post-survey and we interviewed them about their experience using the system, advantages \& disadvantages of the system, and their previous experiences using other editing systems compared to using \sysname{}. The study lasted 100 minutes, and each participant was compensated with 50,000 KRW (approximately 40 USD).

\subsection{Collected Data}


The post-survey contained 7-point Likert-scale questions about participants' confidence in using the system and the perceived helpfulness of the main features of the system. Additionally we also inquiry participants about their experience with the AI tool \cite{wu_ai_2022} and more general System Usability Scale (SUS) \cite{sauro_practical_2011}. We also rated \sysname{} in terms of CSI \cite{cherry_quantifying_2014} for editing informational videos by novice editors. We did not use paired-factor comparison but rather measured the scores from 7-point Likert-scale responses to the questions. We also skipped the ``Collaboration'' scale as our system is not designed with collaboration features in mind. Finally, we ask about the perceived load of the task with the NASA-TLX survey \cite{hart_development_1988}.

To analyze the workflows of the participants while editing an informational video, we logged participants' interactions with the system. We labeled each time segment between interaction logs based on stages of the video editing process that \sysname{} supports (i.e., ideating, describing, examining, and manual editing). Namely, ``ideating'' represents time segments where users were switching between edit layers in ``Edit List'' and history points, which may suggest that participants are actively ideating on what edit to request or implement. The ``describing'' label was given to segments where we captured interactions with the ``Edit Description'' panel (e.g., describing the command with NL, sketching on top of the frame). The ``examining'' segments occur after the user receives the processing results of their multimodal command and starts examining whether to accept or reject the suggestions. Finally, the ``manual editing'' label represents segments where the user actively manipulates the edits by using the ``Canvas'', ``Timeline'', ``Transcript'', or the ``Parameters'' panel.

We decided not to evaluate the outcome videos, as we did not expect participants to complete the task within the given time limit. Rather, our analysis focused more on the video editing process and the experience of the participants using \sysname{}.

\begin{table}[ht!]
    \centering
    \resizebox{\columnwidth}{!}{
        \begin{tabular}{c|c|c|c} \hline 
            Participant &
            Years of Experience &
            Edited videos \#  &
            Assigned video \\ \hline 
            
            P1 & 0 & 2 & EV1\\ \hline 
            P2 & 1 & 6 & EV1\\ \hline 
            P3 & 1 & 3 & EV1\\ \hline 
            P4 & 3 & 5 & EV1\\ \hline 
            P5 & 2 & 2 & EV1\\ \hline 
            P6 & 5 & 5 & EV2\\ \hline 
            P7 & 2 & 5 & EV2\\ \hline 
            P8 & 3 & 4 & EV2\\ \hline 
            P9 & 1 & 15 & EV2\\ \hline 
            P10 & 1 & 1 & EV2\\ \hline
        \end{tabular}
    }
    \caption{The table shows the information about user evaluation participants including the number of years of experience, the reported number of videos they edited, and the assigned video for the study.}
    \Description{The table shows the information (i.e., number of years of experience, the reported number of videos they edited, and the assigned video for the study) about user evaluation participants. There were ten novice participants in total with a diverse number of years of experience. Most of the participants edited at most 5 videos and there were two who edited 6 and 16 videos. The two footage videos were assigned randomly.}
    \label{tab:eval_participants}
\end{table}

\begin{table*}[ht!]
    \centering
    \resizebox{\textwidth}{!}{%
        \begin{tabular}{c|c|c|c|c} \hline 
            Video & Knowledge Characteristic & Content Format & Title & Reference Video\\ \hline 
            EV1 &
            procedural &
            visual \& verbal &
            Jamie Oliver live - pasta \cite{oliver_1} &
            Learn To Cook In Less Than 1 Hour \cite{digiovanni_1} \\ \hline 
            EV2 &
            conceptual &
            verbal &
            How To SURVIVE As An Entrepreneur \cite{perkins_1} &
            the mindset shift that will finally change your work-life \cite{perkins_2} \\ \hline 
        \end{tabular}
    }
    \caption{The table shows the information for the footage videos selected for the user evaluation including knowledge characteristics (procedural or conceptual), content formats (visual or verbal), and video links with respective reference video links.}
    \Description{The table presents the descriptions of the footage videos used in the user evaluation. There are two videos with diverse combinations of knowledge characteristics (procedural or conceptual) and content formats (visual or verbal). Additionally, the table mentions the video links with respective reference video links.}
    \label{tab:eval_videos}
\end{table*}

\section{Results}

Below, we present the results to answer the research questions derived from the post-survey, interview, and observations taken during the study. The post-survey questions and respective results can be found in Appendix~\ref{apx:user_eval_tables}.

\subsection{RQ-1: Editing Patterns and Workflows with \sysname{}}
\subsubsection{Usage Patterns}
All participants started describing their multimodal command with NL and followed it up with the sketch only when they thought it was necessary, which accounted for 25.98\% (STD=22.45\%) of the commands. Many participants performed bulk editing, which applies multiple edits at once by referring to multiple moments within the video (P2, P3, P9, P6, P5), either by using general descriptions (e.g., ``whenever she uses her hand gestures'') or timestamps of the set of moments (e.g., ``add an image of a lamp at the following frames: 9:04, 14:57, 15:22, 15:43, 18:07''). The participants also iterated on their commands by refining and adjusting them when needed (P3, P4, P5, P7, P8, P6, P10). The `Examine' feature especially facilitated the understanding of how the system parsed the commands (P1, P2, P3, P4, P5, P9, P10) (M=6.1/7, STD=0.74) (Table~\ref{tab:nls_usefulness}) and gave rationale behind each suggested edit which helped to decide whether to accept or reject them (P6).

On average, the participants created 5.2 (STD=1.62) multimodal commands and made 9.3 (STD=2.54) requests which included iterations on the commands. In total, 16.6 (STD=6.7) edits are applied to different segments of the video from the commands. 
They accepted 45.98\% (STD=24.8\%) of the suggested edits, and 58.09\% (STD=22.55\%) of their final applied edits were adjusted versions of those edits. The summary of the \sysname{}'s usage is shown in Table \ref{tab:logs_numbers}.

\subsubsection{Usage Workflows}\label{sec:workflows}
There were notable usage patterns of \sysname{} identified during the study. The participants first (1) \textit{ideated} about what edits to implement, (2) \textit{described} their edit intents, (3) \textit{examined} the results returned from the system, and (4) \textit{manually edited} the video.

\begin{figure*}[ht!] 
    \centering
    \includegraphics[width=.8\textwidth]{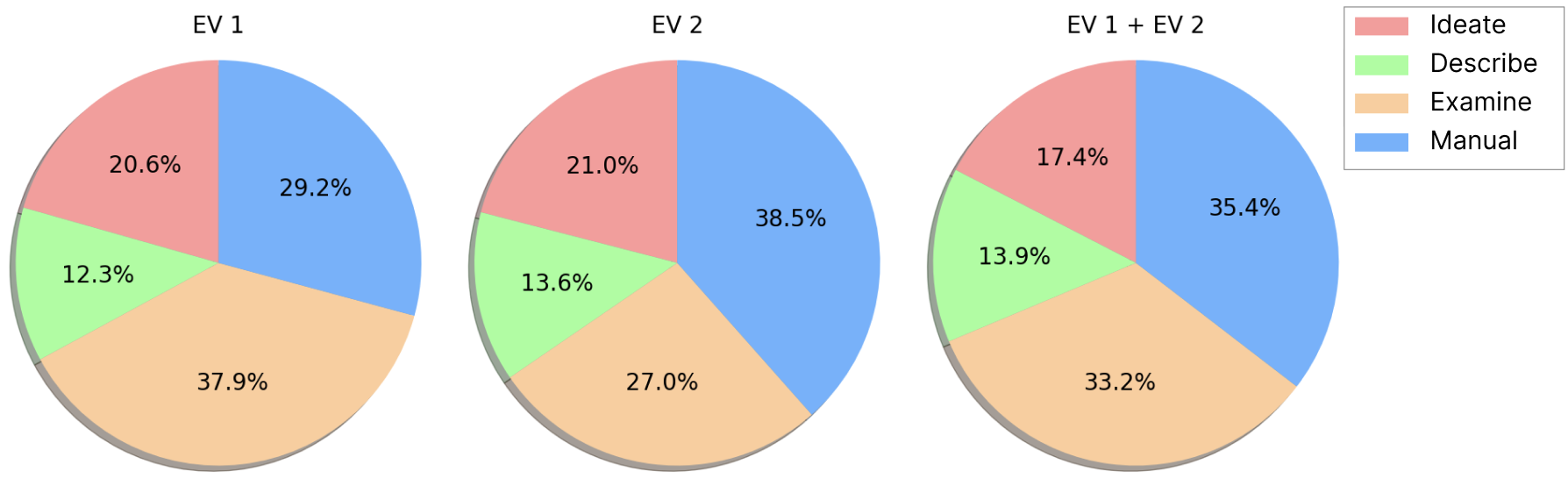}
    \caption{The pie plot shows the distribution of time spent in (1) \textit{ideating} about what edits to implement, (2) \textit{describing} their edit requests, (3) \textit{examining} the suggested edits returned by the system, and (4) \textit{manually editing} the video. The time spent for each editing process stage is derived from user interaction logs.}
    \Description{The pie plot presents the distribution of time spent during the creative process. It is divided into four types: ideating, describing, examining, and manually editing. The first pie plot shows the distribution of the creative process for the first footage video where participants spent 20.6\% of time ideating, 12.3\% of time describing their intents in the form of edit commands, 37.9\% of time examining the generated results from the system, and 29.2\% of the time manually editing. On the other hand, the second plot of the second footage video shows that people spend similar amounts of time ideating and describing edits, but spend 27.0\% of time examining and 38.5\% of time manually editing. This contrast shows how participants show different behaviors based on the complexity of the video (visually complex or verbal). The last pie plot shows the aggregated results for both videos.}
    \label{fig:pie_plot_low}
\end{figure*}

The participants spent a roughly equal amount of time ideating and describing (M=13.9+17.4=31.3\%), examining (M=33.2\%), and manually editing (M=35.4\%) with \sysname{}. Interestingly, different behaviors emerged between the two videos, EV1 and EV2. 
The novices assigned to EV1 mostly examined (M=37.9\%) the processed results of the commands while spending a relatively small amount of time manually editing (M=29.2\%). On the other hand, the novice participants assigned to EV2 spent more time manually applying the edits (M=38.5\%) as opposed to examining the processing results (M=27.0\%). 
This can be explained by the characteristics of the videos. Since EV1 is more visually complex, users need more time to examine the suggested edits by playing and rewinding the part of the video when the edit is applied multiple times to ensure that it fits well. However, in EV2, users could relatively easily decide whether to accept or reject the edit by looking at the transcript or just seeing how the edit looks on the single frame, as the video was somewhat static for the most part. We summarize the aggregated average time spent by participants (for each assigned video) on the editing stages with the pie plot (Figure~\ref{fig:pie_plot_low}). 

\subsection{RQ-2: Understanding and Implementation of Edit Commands}


\subsubsection{Participants thought that \sysname{} understood their commands well}


Overall, the participants felt that the system understood their commands (P6, P3, P8, P4, P5, and P9), especially when the commands were detailed and specific (P6, P1, P2) or when they were based on the transcript or the visual content of the video (P4). At the same time, several participants acknowledged that the system had difficulty understanding broader or less detailed commands (P2, P5).

This aligns with the mixed responses we got from the post-survey regarding how well the system understood the commands (M=4.5/7, STD=1.1) (Table~\ref{tab:nls_performance}). P6, P1, P2, P9, and P10 thought that the misunderstanding might have happened due to their inadequate description of the command. Moreover, at the beginning of the task, a few participants were not sure about what kind of commands they could give (P7, P8) to the system, so they experimented with different commands to test how well the system understands. Fortunately, they said they were able to express their edits more effectively towards the end of the study as they got used to the system more.

\subsubsection{Participants were satisfied with the \sysname{}’s implementation of their commands but adjusted the results to match the video better.}

The participants were generally satisfied with the implementation of the commands (P6, P2, P3, P5, P9, P10) and noted that they got better results when they gave \textit{better} commands (P1, P7), although most of the system-suggested edits were modified or adjusted based on the participant’s preference on how the edit fits the video (i.e., timing, duration, edit parameters). The post-survey results show that the \sysname{}’s implementation of the commands is indeed useful in terms of addressing the implementation of the commands (M=4.5/7, STD=1.4) (Table~\ref{tab:nls_performance}), satisfaction with the quality of the output (M=5/7, STD=1.5) and the performance (M=5.5/7, STD=1.1) (Table~\ref{tab:nls_performance}). P6 and P3 mentioned that they usually got more suggested edits than they were expecting, which was generally viewed as more favorable compared to not getting enough suggestions. Interestingly, P4 said that in certain cases, it was more efficient to manually create edits by using suggestions as markers of important moments in the video, especially for cutting out long segments of the video that cover multiple suggestions.



\subsection{RQ-3: Usefulness of \sysname{}}


\subsubsection{\sysname{} improved the efficiency of editing videos.}

Consistent with our survey results, the participants felt that \sysname{} enhanced their ability to communicate and implement edits (P1, P6, P3, P5, P9) (M=5.6/7, STD=1.17), and the system was easy to use for editing videos (P2, P3, P5, P8, P9) (M=5.6/7, STD=1.17) (Table~\ref{tab:nls_usefulness}). 
Almost all the participants found that they were able to build upon the edits suggested by the system (P1, P2, P4, P5, P6, P7, P8, P9). Specifically, it helped with locating the parts of the video where the edit should be applied (P8, P6, P9). 

The participants also appreciated the sketching on top of the frame, which was mainly used to specify the location of the edit in the frame (P2, P4, P6, P7, P8, P9). P8 and P9 mentioned that it was easier to sketch than manually adjusting the position of the edit (e.g., textbox, image). P4 also acknowledged the convenience of sketching and styling with the command, which reduced the manual work when creating multiple edits. 

Moreover, P6, P7, and P8 noted that describing the command and looking through the suggested edits inspired them and gave more ideas on what edits they could apply. This also aligns well with survey results that shows that the system helped participants think through edits (M=5.7/7, STD=1.42), feel in control of the system (M=6.1/7, STD=0.74), and collaborate with it (M=5.7/7, STD=1.34). The questions and the summary of scores are listed in Table~\ref{tab:aichains}

Despite its advantages, several participants noted that it might be difficult to do more complex and smooth edits (P2) with \sysname{} and that it could miss some moments in the video where they wanted to apply an edit (P4). 

\subsubsection{Comparison with Previous Experiences}

When asked about how the editing experience with \sysname{} differed from previous experiences, many participants noted that the system was much simpler and easier to use. They also mentioned that the system focused on all the basic functionalities to edit an informational video (P1, P3, P5, P9), unlike Adobe Premiere Pro, which has sophisticated but complicated instruments that might confuse novices (P2, P8). However, the more experienced participants noted that \sysname{} lacked some important editing functionalities to feel fully comfortable editing with the system (P1, P3, P4, P5, P9, P6, P8, P10). In terms of workflow, participants noted that they focused more on what edit to apply rather than sticking to the more common chronological order of editing (e.g., adding edits in chronological order of the video) (P4, P5, P9). Moreover, some participants appreciated that, unlike in other video editing tools, they were not starting from scratch but the system was making the initial edits (P1, P3, P5, P9). P8 felt more creative with the \sysname{}'s suggested edits, \emph{``It made my editing process more creative.''}

\subsubsection{Overall feedback on \sysname{}}
The participants felt confident in their ability to use the system (M=5.5/7, STD=1.08), to plan and implement edits (M=5.4/7, STD=0.84), and to revisit and refine edits to improve the quality of the video (M=5.7/7, STD=1.25). However, they had mixed responses about editing different types of videos other than informational videos (M=4.5/7, STD=1.27), producing a quality video (M=4.5/7, STD=0.97), and translating the edits that they envisioned (M=4.7/7, STD=1.16). During the interviews, participants noted that the main reason was the limited editing functionality of \sysname{} and that the quality videos should contain such as smooth animations, transitions, and engaging audio (P2, P9, P6, P1). The summary of the self-confidence scores is listed in Table~\ref{tab:confidence}.
Overall, the system was easy to use, with the SUS usability score of 75.7 (STD=10.00). CSI results show that the system engaged the participants and allowed them to explore multiple options (Table~\ref{tab:csi}). In terms of NASA-TLX, the participants felt moderate mental demand (M=4/7, STD=1.49) and put some effort into doing the tasks (M=3.8/7, STD=1.81), but relatively low workload for the other factors (Table~\ref{tab:nasa_tlx}).
\begin{table*}[ht!]
\centering
\resizebox{.9\textwidth}{!}{%
\begin{tabular}{c|c|c|c|c|c|c|c|c}
\hline
    \multirow{2}{*}{Participant} &
    \multirow{2}{*}{Requests \#} &
    \multirow{2}{*}{\shortstack{Requests\\with Sketch \%}} &
    \multirow{2}{*}{\shortstack{Edit\\Commands \#}} &
    \multirow{2}{*}{\shortstack{Avg. Command\\Iterations \#}} &
    \multirow{2}{*}{\shortstack{Suggested\\Edits \#}} &
    \multirow{2}{*}{\shortstack{Accepted\\Suggestions \%}} &
    \multirow{2}{*}{\shortstack{Applied\\Edits \#}} &
    \multirow{2}{*}{\shortstack{Applied\\Suggestions \%}} \\ & & & & & & & &  \\ \hline
P1    & 10 & 0.0\%  & 5  & 1.00 & 38  & 39.5\% & 18  & 66.7\%  \\ \hline
P2    & 10 & 40.0\% & 6  & 0.67 & 30  & 56.7\% & 18  & 83.3\%  \\ \hline
P3    & 5  & 0.0\%  & 5  & 0.00 & 12  & 83.3\% & 19  & 26.3\%  \\ \hline
P4    & 10 & 20.0\% & 4  & 1.50 & 40  & 15.0\% & 11  & 45.5\%  \\ \hline
P5    & 14 & 14.3\% & 8  & 0.75 & 60  & 28.3\% & 16  & 56.3\%  \\ \hline
P6    & 8  & 37.5\% & 3  & 1.67 & 19  & 31.6\% & 13  & 46.2\%  \\ \hline
P7    & 8  & 75.0\% & 5  & 0.60 & 22  & 72.7\% & 32  & 31.3\%  \\ \hline
P8    & 12 & 33.3\% & 7  & 0.71 & 55  & 29.1\% & 16  & 100.0\% \\ \hline
P9    & 7  & 28.6\% & 6  & 0.17 & 14  & 78.6\% & 17  & 58.8\%  \\ \hline
P10   & 9  & 11.1\% & 3  & 2.00 & 16  & 25.0\% & 6   & 66.7\%  \\ \hline
Total & 93 & -     & 52 & -    & 306 & -     & 166 & -      \\ \hline
Avg. (STD) &
  9.3 (2.54) &
  26.0\% (22.45) &
  5.2 (1.62) &
  0.91 (0.64) &
  30.6 (17.15) &
  45.98\% (24.8) &
  16.6 (6.70) &
  58.1\% (22.55) \\ \hline
\end{tabular}%
}
\caption{The table shows the qualitative summary of the work done by each participant in the user evaluation. The number of processing requests, the percentage of requests with a sketch, the number of individual edit commands, the average number of iterations on those commands, the number of suggested edits by the system for the session, the percentage of accepted suggested edits, the number of final applied edits, and the percentage of initially suggested edits among them.}
\Description{The table contains the qualitative summary of the work done by participants during the user evaluation. The participants made 9.3 requests (i.e., counting both commands from scratch and additional requests) on average and 26\% of them had sketches with it. On average, there were 5.2 individual edit commands and each one was iterated on 0.91 times. In total, novices received 30.6 edits as a suggestion from the system and accepted almost 50\% of them. In the final version of the video, there were 16.6 edits applied on average, and among them, 58\% were system-suggested.}
\Description[short]{long}
\label{tab:logs_numbers}
\end{table*}

\section{Discussion}

In this paper, we present \sysname{}, a multimodal system for editing informational videos. It is powered by CV and LLM-based pipeline that supports multimodal reasoning and real-time processing of natural language and sketching (NL\&S) commands. We discuss the importance of balancing expressiveness and control, how \sysname{} supports the creative process, how video characteristics affect the system usage, considerations for utilizing AI for video editing, and limitations and future work.

\subsection{Balancing Expressiveness and Control}

With \sysname{}, we observed that the participants were able to edit videos efficiently and in a creative way when the system appropriately restricts the NL command usage to express the video editing requests to three defining aspects of a video edit; temporal location, spatial location, and edit operation and parameters. The participants gradually figured out what kind of commands the system supports and effectively used them in a relatively short duration of our study. Moreover, the three types of references facilitated the iteration of the edit commands and helped users detect where the system misinterpreted the command and fix it. Thus, we highlight the importance of balancing the sense of control that users have and the expressiveness that the system supports. While more advanced language models can allow video editing systems to support a broader set of NL expressions, the trade-off between expressiveness and control should be considered.

\subsection{Supporting Different Phases of the Creative Process}

The creative process consists of three iterative phases: ideation, execution, and evaluation \cite{candy_evaluating_2013}. \sysname{} mainly supports the execution stage by allowing users to express and implement their edits more naturally and efficiently based on multimodal edit commands. However, our system can also help with the ideation and evaluation phases. For instance, users can describe the text content that they want on the video, and the system can suggest moments in the video and spatial location in a frame where such content could appear. It can help users judge if such an edit idea is appropriate and iterate on the idea at a low cost, which can then serve as a starting point for users to work on. Moreover, \sysname{} can be useful at the evaluation stage of the creative process. It helps users evaluate if an edit would be appropriate or look \textit{good} by automatically generating the edits suitable for each context (e.g., the appropriate spatial location and parameters). As such, \sysname{} can support each of the creative process phases in video editing through multimodality.

\subsection{Impact of Video Characteristics}
From the user study, we observed that the participants exhibited varying editing behaviors when working with different videos, possibly due to the distinct characteristics of the videos --- Participants with visually complex videos spent more time examining the processed results and less time manually editing, while the participants with verbally-oriented videos demonstrated the converse pattern of behavior (Section~\ref{sec:workflows}). It suggests that the workflow and design of the system can be improved in a way that reflects the video characteristics, such as displaying the processed results in an easy-to-skim manner with visually complex videos. Furthermore, types of edit operations that are effective or frequently used could be different depending on the knowledge characteristic of the video content, such as procedural (e.g., cooking videos) and conceptual (e.g., mathematics lecture videos). Creating a system tailored to a particular video type can enhance the user experience by providing relevant editing suggestions.

%

\subsection{Using AI for Video Editing}

When developing AI-infused video editing tools for novices, it is crucial to acknowledge the ethical considerations such as inappropriate use of the system and its broader impact on novice users.

As existing work has shown \cite{jakesch_co-writing_2023, cao_comprehensive_2023}, AI models can generate inappropriate output that can have significant consequences when infused into interactive systems such as ours. For example, our CV and LLM-based pipeline can potentially generate edits with harmful or hallucinated content when editing educational videos. We follow common approaches to address these issues \cite{amershi_guidelines_2019, lipton_mythos_2017} by providing a rationale behind each generated output of our pipeline and allowing users to manually revise the edits. Nevertheless, we emphasize the importance of addressing AI bias and ethics when developing AI-infused video editing tools.

As we observed in our study, \sysname{} can effectively support novice video editors in creating video edits. While it is important to lower the barrier to engaging with video editing tools, it is equally important to ensure that the system allows novices to improve their skills and confidence in video editing. Thus, we acknowledge the potential of users' overreliance on AI-based interpretation of the edit commands and reduced self-confidence in video editing \cite{bucinca_trust_2021, kim_stylette_2022}. We partly address this aspect by enabling manual video editing in \sysname{}, however, additional features within the system that prompt the users to reflect on the generated edits \cite{shin_understanding_2018} can promote better deliberate learning and understanding of video editing practice.

\subsection{Limitations and Future Work}
Although our study revealed the usefulness of \sysname{} in video editing, there are a few limitations of our system and the study. In the formative study, we focused on edit commands that initiate edits rather than revise or adjust applied edits. We believe future work can build on top of our initial investigation . In \sysname{}, the pipeline handles interpreting three reference types in an NL command. While these types appeared the most in our formative study, we also observed other uses of NL by participants, such as describing edit rationale and intended effect on the audience. Furthermore, the system supports sketching on top of the frame mainly to indicate the region of interest within the single frame. As it was observed in our study, there could be other cases where sketching is used, such as sketching the content that should be added or indicating the movement by sketching on multiple keyframes. Lastly, participants felt limited by the seven edit operations, which were chosen based on their frequency in the formative study. Thus, future work can investigate improving the pipeline to include a more diverse range of edit intents, and to extend the role of sketching and the set of supported edit operations beyond visual effects such as animation and audio manipulation.



We conducted an observational study with 10 novice video editors to evaluate how effective the pipeline is and how useful the system is. Future work can also evaluate \sysname{} with a deployment study or in a more comparative setting to get more statistical insights into the benefits of the system. Moreover, the participants worked on pre-selected footage videos as opposed to their own videos. While this scenario is close to real-world tasks that video editors encounter, our participants had difficulties becoming familiar with the content of the videos and generating edit ideas within the duration of the study. To eliminate such challenges, \sysname{} can be evaluated with the participants' videos, similar to the evaluation by Huh et al. \cite{huh_avscript_2023}.



%










\section{Conclusion}
We propose \sysname{}, a multimodal video editing system that allows users to edit videos using natural language (NL) and sketching on top of a video frame. Based on findings from the formative study and the analysis of 176 multimodal edit commands, our technical pipeline powered by CV and large language models is designed in a way that it extracts and interprets (1) temporal, (2) spatial, and (3) operational references in an NL command and spatial references from sketching. Our system implements the interpreted edits which then users can iterate on. Our study with 10 participants shows that \sysname{} helps users generate and express edit ideas and implement them effectively. We believe that our work opens up new opportunities in natural language video editing and multimodal interfaces.

\begin{acks}
This work was supported by Institute of Information \& Communications Technology Planning \& Evaluation (IITP) grant funded by the Korea government (MSIT) (No.2021-0-01347,Video Interaction Technologies Using Object-Oriented Video Modeling).
\end{acks}

\bibliographystyle{ACM-Reference-Format}
\bibliography{references-zotero, references}

\onecolumn
\clearpage
\def\lstlistingname{Prompt}
\appendix
\section{Ground Truth Construction for the Technical Pipeline} \label{apx:gt}

When constructing the ground truth dataset to evaluate the technical pipeline, the following procedure was used to analyze each of the 50 edit commands:
\begin{enumerate}
    \item if the edit command contained a screenshot of the frame with the timestamp and a clear indication of where the edit should be within the frame, then we used that information directly within the edit command's spatial and temporal parameters. 
    \item if the edit command did not contain a screenshot of the frame, then we analyzed the natural language part of the command. For references that mentioned explicit timestamps (e.g., "at 15:57,...") or absolute positions within the video frame (e.g., "top left corner"), we used that information directly within the edit command's spatial and temporal parameters.
    \item For NL commands with more conditional language (e.g., "whenever the camera is facing down to the pan"), we manually identified parts of the video that satisfied these conditions. For edit commands with vague or non-existent spatial references, we placed the edits in the frame such that it did not block the important content in the video (e.g., speakers, objects that they are interacting with). 
\end{enumerate}



\section{User Evaluation Results} \label{apx:user_eval_tables}
\begin{table}[ht!]
    \centering
    \begin{tabular}{p{0.65\columnwidth}|c}
    \hline
    Criteria & Avg. (STD) \\ \hline
    1. I feel confident in my ability to navigate and use the system. & 5.5 (1.08)\\ \hline
    2. I am confident in my skills in editing various types of videos (visual/verbal) using this system. & 4.5 (1.27) \\ \hline
    3. I am confident that I can produce quality videos using this system. & 4.5 (0.97) \\ \hline
    4. I am confident that I can plan edits with the system and implement them. & 5.4 (0.84) \\ \hline
    5. I am confident that I can translate what I envision into actual video edits with this system. & 4.7 (1.16) \\ \hline
    6. I am confident that I can revisit/refine my video edits to achieve better results with this system. & 5.7 (1.25) \\ \hline
    \end{tabular}
    \caption{The table shows the survey results for self-confidence ratings of the participants from the user evaluation in terms of a 7-point Likert-scale.}
    \Description{The table presents the 7-point Likert-scale post-survey results regarding the self-confidence of the participants in terms of how confident they are in navigating, editing different types of videos, producing quality videos, planning \& implementing edits, translating envisioned edits into actual video, and refining the edits. In general, the results show that the novices felt confident in their ability to use the system and refine their edits. However, there were mixed responses regarding editing \& producing different types of quality videos and translating the edits that they envisioned into actual edits.}
    \label{tab:confidence}
\end{table}

\begin{table}[ht!]
    \centering
    \begin{tabular}{p{0.65\columnwidth}|c}
    \hline
    Criteria & Avg. (STD) \\ \hline
    1. The system understood my edit commands (in natural language and sketched form) well. & 4.5 (1.08) \\ \hline
    2. The system implemented my edit commands (in natural language and sketched form) well. & 4.5 (1.43) \\ \hline
    3. I am satisfied with the outcome quality of the edit command (natural language request and sketching) processing. & 5 (1.49) \\ \hline
    4. I am satisfied with the performance of the edit command (natural language request and sketching) processing. & 5.5 (1.08) \\ \hline
    
    \end{tabular}
    \caption{The table shows the survey results for questions about the perceived performance of the \sysname{}'s processing in terms of a 7-point Likert-scale.}
    \Description{The table presents the post-survey results from the user evaluation for 7-point Likert-scale questions regarding the performance of the pipeline. 
    The participants had mixed responses regarding how well the system understood them. They thought that the misunderstanding might have happened due to their inadequate description of the command. Also, most participants tried experimenting with the system to explore its capabilities and said that they were able to use the system more effectively toward the end of the study. Moreover, the results show that users were satisfied with the edit command implementation of the system).}
    \label{tab:nls_performance}
\end{table}

\begin{table}[ht!]
    \centering
    \begin{tabular}{p{0.65\columnwidth}|c}
    \hline
    Criteria & Avg. (STD) \\ \hline
    1. I am satisfied with my final results from the system; they met the task goal. & 3.9 (1.66) \\ \hline
    2. The system helped me think through what kinds of output I would want to complete the task goal, and how to complete the task. & 5.7 (1.42) \\ \hline
    3. The system is transparent about how it arrives at its final result; I could roughly track its progress. & 5.6 (1.35) \\ \hline
    4. I felt I had control creating with the system. I can steer the system towards the task goal. & 6.1 (0.74) \\ \hline
    5. In the system, I felt I was collaborating with the system to come up with the outputs. & 5.7 (1.34) \\ \hline
    
    \end{tabular}
    \caption{The table shows the survey results for self-perceived experience with AI tool \cite{wu_ai_2022} in terms of a 7-point Likert-scale.}
    \Description{The table presents survey results for questions regarding self-perceived experience with AI tools based on a 7-point Likert-scale. They thought that the system helped participants think through edits they wanted to make, feel in control of the system, and collaborate with it.}
    \label{tab:aichains}
\end{table}

\begin{table}[ht!]
    \centering
    \begin{tabular}{p{0.65\columnwidth}|c}
    \hline
    Criteria & Avg. (STD) \\ \hline
    1. It was easy to understand and use the Edit description (natural language request and sketching) features in the system. & 5.6 (1.17) \\ \hline
    2. The Edit description (natural language request and sketching) features enhanced my ability to communicate and implement video edits. & 5.6 (1.17) \\ \hline
    3. The Examine (breakdown of the natural language request) feature helped me understand how the Edit description feature (natural language request and sketching) works. & 6.3 (0.48) \\ \hline
    4. The Examine (breakdown of the natural language description) feature helped me iterate (refine/adjust) the Edit descriptions (natural language request and sketching) I made. & 6.1 (0.74) \\ \hline
    
    \end{tabular}
    \caption{The table shows the survey results for questions about the usefulness of the ``Edit Description`` and ``Examine`` features of \sysname{} in terms of a 7-point Likert-scale.}
    \Description{The table presents the 7-point Likert-scale post-survey results regarding the usefulness of Edit Description and Examine features in the system. The table results show that participants felt that the Edit description feature was easy to understand and enhanced their expression of video editing intents. Moreover, the Examine feature helped them more easily understand the rationale behind system-suggested edits.}
    \label{tab:nls_usefulness}
\end{table}

\begin{table}[ht!]
    \centering
    \begin{tabular}{c|c}
    \hline
    Criteria & Avg. (STD) \\ \hline
    Enjoyment & 6.1 (1.02) \\ \hline
    Exploration & 6.1 (1.05) \\ \hline
    Expressiveness & 4.8 (1.36) \\ \hline
    Immersion & 4.6 (1.67) \\ \hline
    Results Worth Effort & 5.0 (1.59) \\ \hline
    \end{tabular}
    \caption{The table shows the Creativity Support Index scores for Enjoyment, Exploration, Expressiveness, Immersion, and Results Worth Effort in terms of a 7-point Likert-scale.}
    \Description{The table presents Creativity Support Index scores for Enjoyment, Exploration, Expressiveness, Immersion, and Results Worth Effort from the user evaluation based on a 7-point Likert-scale. It shows that the participants enjoyed working with the system and it allowed them to explore different edit ideas. However, they felt that the system could have supported better expressiveness of natural language and immersion. Lastly, they had mixed opinions regarding whether their effort was worth the result.}
    \label{tab:csi}
\end{table}

\begin{table}[ht!]
    \centering
    \begin{tabular}{c|c|c|c|c|c|c}
    \hline
    \multicolumn{1}{l|}{Video} &
      \multicolumn{1}{l|}{Mental} &
      \multicolumn{1}{l|}{Physical} &
      \multicolumn{1}{l|}{Temporal} &
      \multicolumn{1}{l|}{Effort} &
      \multicolumn{1}{l|}{Performance} &
      \multicolumn{1}{l}{Frustration} \\ \hline
    EV1   & 4 (1.58) & 2.4 (1.14) & 4.4 (1.52) & 3.8 (2.17) & 2.4 (0.55) & 1.6 (0.89) \\ \hline
    EV2   & 4 (1.58) & 1.4 (0.55) & 3.4 (2.89) & 3.8 (1.64) & 4.4 (1.52) & 1.8 (1.79) \\ \hline
    Total & 4 (1.49) & 1.9 (0.99) & 3.9 (2.23) & 3.8 (1.81) & 3.4 (1.51) & 1.7 (1.34) \\ \hline
    \end{tabular}%
    \caption{The table shows the reported NASA-TLX scores (i.e., average (std)) from the user evaluation in terms of a 7-point Likert-scale.}
    \Description{The table presents the reported NASA-TLX scores from the user evaluation in terms of the 7-point Likert-scale. Most notably, the participants felt moderate mental demand and put some effort into doing the tasks but felt a relatively low workload for the other factors. }
    \label{tab:nasa_tlx}
\end{table}

\clearpage

\section{Prompts used as part of the Technical Pipeline} \label{apx:prompts}
\subsection{Parsing Edit Command} \label{apx:prompt_1}

\begin{lstlisting}[breaklines=true, breakatwhitespace=true, basicstyle=\ttfamily\small, captionpos=b, caption=\textbf{The prompt for Stage 1.}]
You are a video editor's assistant who is trying to understand the natural language command in the context of a given video. You will do it step-by-step.

Step 1: Identify the list of edit operations that the command is referring to:
- choose only among "text", "image", "shape", "blur", "cut", "crop", "zoom"
- make sure that the edit operation is only one of the above
- if none of the above edit operations is directly relevant, give the one that is most relevant to the command (e.g. "highlight" -> "shape" with type parameter "star")

Step 2: You have to identify 3 types of references from the command (Note: if there is a reference that contains noun-references such as this, that, it, etc. you will have to identify the noun that it refers to and replace the noun-reference with the noun.):
1. Temporal reference: any information in the command that could refer to a segment of the video:
- explicit timecodes or time ranges
- explicit mentions or implicit references to the transcript of the video
- description of the actions that happen in the video
- visual description of objects, moments, and frames in the video

2. Spatial reference: any information in the command that could refer to location or region in the video frame:
- specific locations or positions relative to the frame
- specific objects or areas of interest

3. Edit Parameter reference: any information in the command that could refer to specific parameters of an edit operation that was identified ([text, image, shape, blur, cut, crop, zoom]).
- text: content, font style, font color, or font size
- image: visual keywords
- shape: type of shape
- blur: degree of blur to apply
- cut: no parameters
- crop: how much to crop
- zoom: how long to perform the zooming animation

Step 3-1: You will classify each temporal reference into one of the following:
1. "position": reference in the form of a timecode (e.g. "54:43", "0:23"), time segment (e.g. "0:00-12:30", "from 43:30 to 44:20") or more abstract temporal position (e.g. "intro", "ending", "beginning part of the video")
2. "transcript": reference to transcript both implicit or explicit
3. "video": reference to specific action in the video or visual description of the frame, object, or elements
4. "other": reference to other temporal information that does not fall into the above categories

Step 3-2: You will classify each spatial reference into one of the following:
1. "visual-dependent": reference to specific objects, elements, or regions in the video frame that depend on the visual content of the video
2. "independent": reference to specific locations or positions relative to the frame independent of the visual content of the video
3. "other": any other spatial information that does not fall into the above categories

Step 4: Format the output based on the result of each step.

{ few-shot examples... }
\end{lstlisting}

\subsection{Temporal Interpretation} \label{apx:prompt_2}

\begin{lstlisting}[breaklines=true, breakatwhitespace=true, basicstyle=\ttfamily\small, captionpos=b, caption=\textbf{The prompt for Stage 2.}]
You are a video editor's assistant who is trying to understand natural language temporal reference in the video. You will do it step-by-step.

First step: Identify the type of temporal reference based on the user's command.
1. Timecode: a specific time in the video
2. Time range: a range of time in the video
3. More high level temporal reference: a reference to a generic event in the video (introduction, ending, etc.)

Second step: Identify the timecode or time range with additional context.
Note 1: If the temporal reference is just a timecode, output any 10 second interval containing the timecode.
Note 2: If there are more than one segment of video that matches the temporal reference, output all of them in a list.

{ few-shot examples... }
\end{lstlisting}

\begin{lstlisting}[breaklines=true, breakatwhitespace=true, basicstyle=\ttfamily\small, captionpos=b, caption=\textbf{The prompt for Stage 2 transcript references.}]
You are a video editor's assistant who is trying to understand the natural language reference of the video editor to some part of the video given the original context of the reference and relevant snippets of the transcript of the video.

Instruction:
Locate the snippets of the transcript that are relevant to the editor's command and original context, and return the positions of those snippets from the list along with short explanation of how each one is relevant to editor's command and original context.

Note 1: If there are no relevant snippets, return an empty array [].
Note 2: If there is more than one snippet that is relevant to the editor's command, output all of them in a list with respective indexes and explanations.

{ few-shot examples... }
\end{lstlisting}

\begin{lstlisting}[breaklines=true, breakatwhitespace=true, basicstyle=\ttfamily\small, captionpos=b, caption=\textbf{The prompt for Stage 2 video references.}]
You are a video editor's assistant who is trying to understand the natural language reference of the video editor to some part of the video given the set of most relevant visual descriptions of 10-second clips of the video and original context of the command. Visual description of a 10-second clip consists of an action label which is a main action happening, an abstract caption which is an abstract description of the clip, and the dense captions, which are list of descriptions of objects that are present. Try taking into account each of them. 

Instruction:
Locate the visual descriptions that are relevant to the editor's command and original context of the command, and return the positions of those descriptions from the list along with short explanation of how each is relevant to editor's command.

Note 1: If there are no relevant viusal descriptions, return an empty array [].
Note 2: If there is more than one description that is relevant to the editor's command and original context, output all of them in a list.

{ few-shot examples... }
\end{lstlisting}

\subsection{Spatial Interpretation} \label{apx:prompt_3}

\begin{lstlisting}[breaklines=true, breakatwhitespace=true, basicstyle=\ttfamily\small, captionpos=b, caption=\textbf{The prompt for Stage 3.}]

You are a video editor's assistant who is trying to understand editor's natural language description of the spatial location within the frame. The description is based on the rectangle that is already present in the frame. You will have to refine its location and resize (if necessary) based on the command.
You will be given the initial location of the rectangle in the frame: x, y, width, height, where (x, y) are coordinates of the top-left corner, and (width, height) are just width and height. Also, you will be given a command that describes the desired spatial location of the rectangle in the frame, the original context of the command, and the boundaries of the frame (e.g. width=1280, height=720)

You will do it step-by-step.
1. Refine the location of the rectangle (x, y coorindates) based on the command, original context of the command, and boundaries of the frame (make sure not to exceed the boundaries);
2. Resize the rectangle (width, height) based on the command, original context of the command, and boundaries of the frame (make sure not to exceed the boundaries);

Perform each step one-by-one and output the final location of the rectangle in the frame in appropriate format.

{ few-shot examples... }
\end{lstlisting}

\subsection{Edit Operation and Parameters Interpretation} \label{apx:prompt_4}

\begin{lstlisting}[breaklines=true, breakatwhitespace=true, basicstyle=\ttfamily\small, captionpos=b, caption=\textbf{The prompt for Stage 4.}]
You are a video editor's assistant who is trying to understand video edit parameter change requests in natural language. You are given a natural language command from the editor, the original context of the command, and initial values of the video edit parameters. You have to appropriately change the parameters to satisfy the command within its original context. You will do it step-by-step.

Step 1: Identify the type of each edit parameter change based on the user's command. There are three types of video edit parameter change requests:
1. Explicit: explicit values for a parameter (e.g. 12px, 10%, "Introduction", etc.)
2. Relative: a relative change to a parameter (e.g. 5 seconds longer, 10% less, fewer words, etc.)
3. Abstract: an abstract change to a parameter (e.g. shorter, longer, more, less, etc.)

Step 2: Transform each type of parameter change request into parameter values based on the "Initial parameters" provided and output the adjusted set of video edit parameters.

{ few-shot examples... }
\end{lstlisting}

\begin{lstlisting}[breaklines=true, breakatwhitespace=true, basicstyle=\ttfamily\small, captionpos=b, caption=\textbf{The prompt for Stage 4 image search query.}]

You are a video editor's assistant who is trying to understand natural language request of the editor to come up with search query for images to put in the video. You are given a command from the editor, the original context of the command, and relevant content from the video. Relevant content is a list of snippets from the transcript and visual description (what action is happening, abstract caption, and descriptions of objects) of 10-second segments. You must generate the search query for the image to be displayed based on the editor's command, original context, and relevant content.

Note 1: If no relevant search query can be generated that satisfies the command, output only the command.
Note 2: Make sure that the search query is not too long, since it should be seen by the editor. Keep it under 100 characters.

{ few-shot examples... }
\end{lstlisting}

\begin{lstlisting}[breaklines=true, breakatwhitespace=true, basicstyle=\ttfamily\small, captionpos=b, caption=\textbf{The prompt for Stage 4 text insertion parameter interpretation.}]

You are a video editor's assistant who is trying to understand natural language request of the editor to find a text to display in the video. You are given a command from the editor, the original context of the command, and relevant content from the video. Relevant content is a list of snippets from the transcript and visual description (what action is happening, abstract caption, and descriptions of objects) of 10-second segments. You must generate the text to be displayed based on the editor's command, original context, and relevant content.

Note 1: If no relevant text can be generated that satisfies the command, output the input command itself with reasonable formatting.
Note 2: Make sure that text is not too long, since it will be displayed on the screen. Keep it under 100 characters.

{ few-shot examples... }
\end{lstlisting}









\end{document}